\documentclass[superscriptaddress,preprintnumbers,prl,twocolumn]{revtex4}
\usepackage{amsmath}%
\usepackage[table]{xcolor}
\usepackage{amsfonts}%
\usepackage{amssymb}%
\usepackage{graphicx}
\usepackage{xcolor}
\usepackage{enumerate} 
\usepackage[english]{babel}
\usepackage{bbold} 
\usepackage[a4paper]{geometry}
\usepackage{multirow}
\usepackage{xcolor}

\usepackage{graphicx}
\usepackage{soul}

\begin{document}

\title{{Manipulating magnetic fields in inaccessible regions} by negative magnetic permeability}

\author{Rosa Mach-Batlle}
\affiliation{Departament de Fisica, Universitat Autonoma de Barcelona, Bellaterra 08193, Spain}
\affiliation{Center for Biomolecular Nanotechnologies, Istituto Italiano di Tecnologia, Via Barsanti 14, 73010 Arnesano (LE), Italy}

\author{Mark G. Bason}
\affiliation{Department of Physics and Astronomy, University of Sussex, Falmer, Brighton BN1 9QH, United Kingdom}

\author{Nuria Del-Valle}
\affiliation{Departament de Fisica, Universitat Autonoma de Barcelona, Bellaterra 08193, Spain}

\author{Jordi Prat-Camps}
\email{j.prat.camps@gmail.com}
\affiliation{INTERACT Lab, School of Engineering and Informatics, University of Sussex, Brighton BN1 9RH, United Kingdom}

\begin{abstract}
The control of magnetic fields underpins a wide variety of cutting-edge technologies, ranging from the guiding of micro-robots for medical applications to the trapping of atoms for quantum experiments. 
Relying on different arrangements of coils, these technologies demand advances in the manipulation of magnetic fields {at a distance from the sources that generate them.}
%
Here, we present a strategy for remotely shaping magnetic fields based on active magnetic metamaterials with negative permeability and we theoretically and experimentally demonstrate the possibility of emulating and cancelling magnetic sources at distance.
{Our strategy provides magnetic field focusing deep into physically inaccessible volumes 
where neither sources of magnetic field nor magnetic materials can be placed.}

\end{abstract}

\maketitle 

Magnetism is a physical phenomenon that is found at the basis of a wide range of technologies, from conventional motors and generators to sophisticated biomedical techniques and quantum technologies. In the last decade, the advent of metamaterials has enabled the manipulation of magnetic fields in unprecedented ways, leading to magnetic cloaks that can make objects magnetically undetectable \cite{Narayana2012a,Gomory2012}, magnetic hoses that can guide the magnetic field \cite{Navau2014}, and even magnetic wormholes that can connect two points through an undetectable path \cite{Prat-Camps2015}.
However, the degree of control one can exert over magnetic fields is restrained by some fundamental limitations. One of the most striking examples is the impossibility of generating local magnetic field maxima in free space, as stated by Earnshaw's theorem~\cite{Wing1984, Ketterle1992}. This strong limitation prevents the stable magnetostatic trapping of paramagnetic and ferromagnetic materials, limits the ability to concentrate magnetic fields, and makes it impossible to exactly create and cancel magnetic field sources, such as magnets or electric wires, at a distance. 
%



Here we propose a strategy for manipulating magnetic fields remotely that makes it possible to overcome these stringent limitations. In particular, we demonstrate the possibility of emulating the field of a magnetic source (a straight current wire) at a distance from the structure that generates such field. Our approach is based on negative-permeability magnetic materials, which have recently been shown to exhibit appealing unconventional properties \cite{Mach-Batlle2018a}. Such materials do not naturally occur \cite{Dolgov1983}, but they can be effectively mimicked as active metamaterials consisting of precise arrangements of currents \cite{Mach-Batlle2017a}. The emulation of magnetic sources remotely is not only scientifically intriguing but could also have deep impacts on several magnetic field-based, cutting-edge technologies. For example, it could improve the guidance of magnetic microrobots and functional nano-particles for medical applications~\cite{Sitti2009,Estelrich2015}, enhance techniques like transcranial magnetic stimulation~\cite{Wagner2007} {by providing strong and spatially-localized magnetic fields at target locations}, open new venues in atom trapping, or provide radical new ways of shielding undesired magnetic fields remotely.

\begin{figure}[h!]
	\centering
	\includegraphics[width=0.45\textwidth]{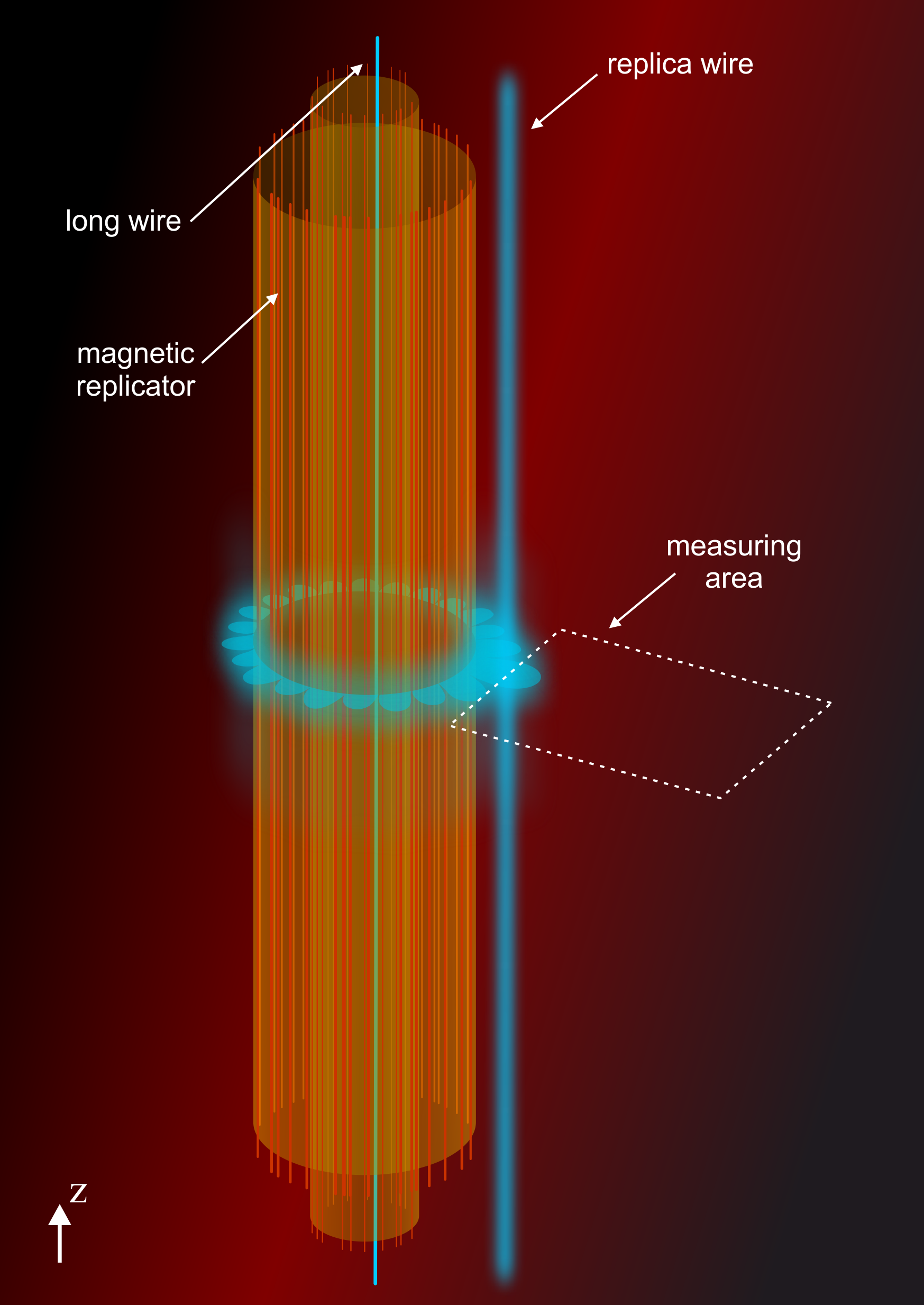} 
	\caption{A {magnetic replicator}. Artistic sketch of the device. A long wire is surrounded by a {magnetic replicator} generating a {replica} of the wire outside it. \label{Fig1}}
\end{figure}

Based on transformation optics (see the Supplementary Material), we demonstrate that a long cylindrical shell with relative magnetic permeability $\mu = -1$ acts analogously to a perfect lens for electromagnetic waves~\cite{Pendry2002,Pendry2003} in the magnetostatic limit. When this negative-permeability shell surrounds a magnetic field source 
the field distribution outside the shell corresponds to the field created by another magnetic source, a {\textit{replica}} source, which can appear at a distance from its external surface, i.e. in {free} space (see Fig. 1). Even though materials with negative permeability do not exist, a cylindrical shell with $\mu = -1$ can be {mimicked} by two surface current densities, ${\bf K}_{\rm M_1}(\varphi)$ and ${\bf K}_{\rm M_2}(\varphi)$, flowing on its internal and external surfaces, respectively~\cite{Mach-Batlle2017a,Mach-Batlle2018a}.
Our goal now is to demonstrate that 
these current densities can lead to the creation of a magnetic source at a distance. For the sake of generality, we focus on the creation of a wire; any other field {source} with translational symmetry along the device's axis can be regarded as the superposition of the field created by a combination of wires. 

We solve the magnetostatic Maxwell's equations considering a long (along $z$) cylindrical shell with $\mu=-1$, internal radius $R_1$ and external radius $R_2$ which we coin a {\textit{magnetic replicator}}. It surrounds a long (along $z$) straight wire of current $I$ placed at $(x,y) = (d,0)$.
The solutions show that the magnetic field in the space beyond the replicator corresponds to the field created by a {replica} wire of current $I$ placed at $(x,y) = (d',0)$, where $d' = d R_2^2/R_1^2$. When the {replica} wire appears in empty space, $d'>R_2$, the field solution in the region $\rho \in (R_2$, $d')$, diverges \cite{AnanthaRamakrishna2004,Milton2005} (where $\rho$ is the standard cylindrical radial coordinate). 
This annular region of divergent field solution makes it possible to have a divergence of the field at the position of the {replica wire} without having a local field maximum in empty space, thus ensuring that Earnshaw's theorem is obeyed (see the Supplementary Material)
%

From the field distribution, we obtain the current densities required to emulate a wire at a distance. Both ${\bf K}_{\rm M_1}(\varphi)$ and ${\bf K}_{\rm M_2}(\varphi)$ are found as an infinite series but, while ${\bf K}_{\rm M_1}(\varphi)$ is always convergent, ${\bf K}_{\rm M_2}(\varphi)$ only converges when the {replica} wire appears in the {device} volume ($d' \leq R_2$). Therefore, the creation of a {replica} wire in empty space ($d'>R_2$) would require infinite currents at the  external boundary of the {device}. However, one can still approximate well the field created by a wire at a distance by truncating the summation in ${\bf K}_{\rm M_2}(\varphi)$ up to $n_{\rm T}$ terms. The higher the number of terms ${n_{\rm T}}$, the more the external field distribution resembles the field created by a wire (compare Figs.~2A-C with D). As shown in Fig.~2E, the spatial focusing of the $B_y$ component is greatly improved by adding more terms. The $B_x$ component also converges to the distribution created by a wire by increasing ${n_{\rm T}}$ (see Fig.~2F).

The inspection of the role of the two current densities ${\bf K}_{\rm M_1}(\varphi)$ and ${\bf K}_{\rm M_2}(\varphi)$ leads to a simplification of our device. The creation of a straight {replica} wire carrying a current $I$ at a distance $d'$ from the shell center only requires a centered wire of current $I$ and the surface current density 
    \begin{equation}
    {\bf K}^{\rm n_T}(\varphi) = \sum_{n=1}^{n_{\rm T}} \frac{I}{\pi R_2}\left(\frac{d'}{R_2}\right)^n{\rm cos}(n \varphi)  {\bf z}
    \end{equation}
fed at the cylindrical surface $\rho = R_2$ (see details in the Supplementary Material). Therefore the information on the explicit values of $d$ and $R_1$ are irrelevant to our experimental implementation.  

\begin{figure}[t]
	\centering
	\includegraphics[width=0.5\textwidth]{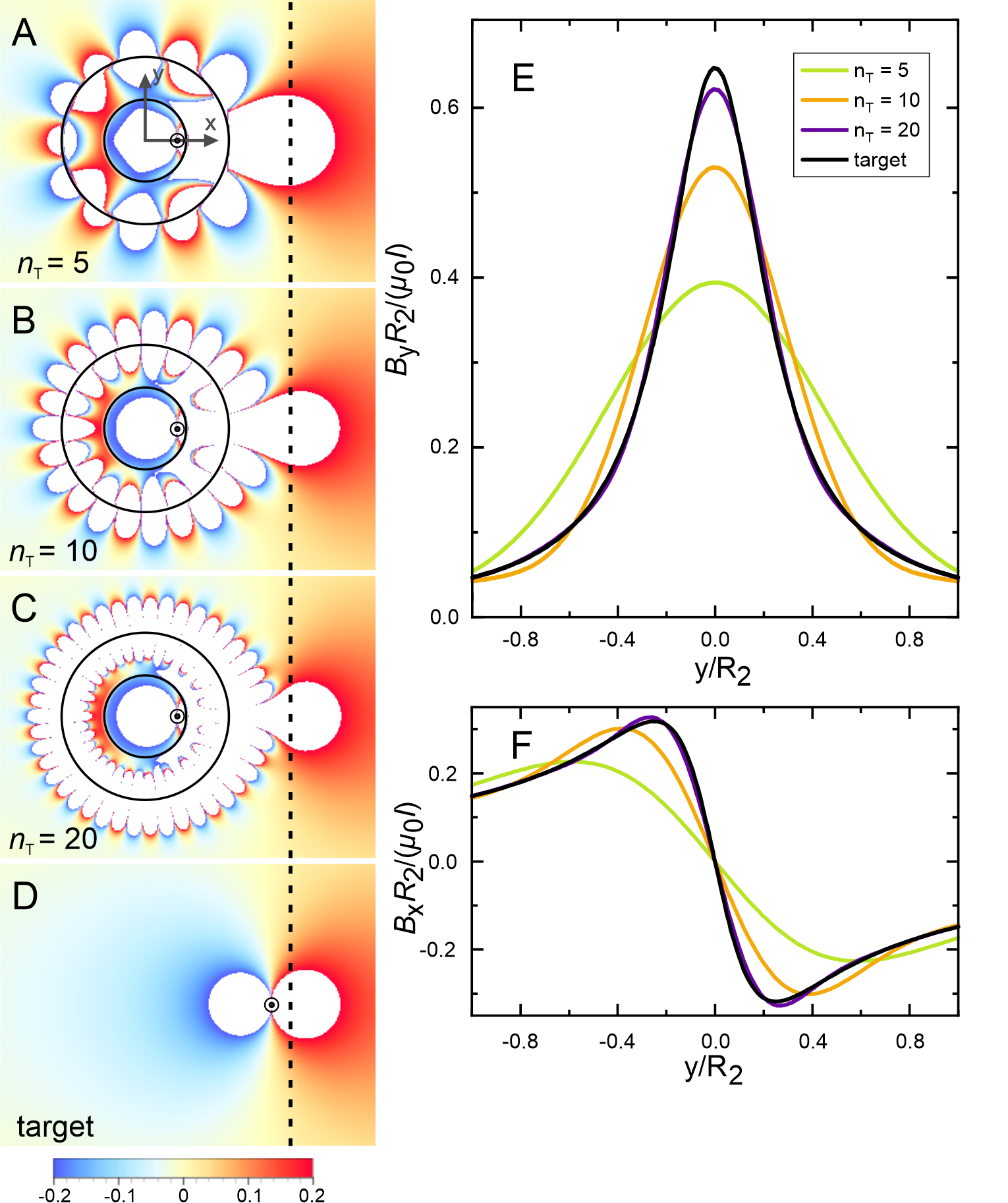} 
	\caption{Numerical calculations of a {magnetic replicator}. (A)-(C) Plots of the normalized $y$-component of the magnetic induction $B_y R_2/(\mu_0 I)$ along the plane $XY$ for a long (along $z$) straight wire carrying a current $I$ placed at $(x,y) = (3/8 R_2,0)$ surrounded by the magnetization currents of a {magnetic replicator} of internal radius $R_1= 1/2 R_2$, external radius $R_2$ and relative magnetic permeability $\mu = -1$ centered at the origin of coordinates. The magnetization current density ${\bf K}_{\rm M_2}$ is truncated to (A) $n_{\rm T} = 5$ terms, (B) $n_{\rm T} = 10$ terms, and (C) $n_{\rm T} = 20$ terms. (D) Calculation of {the field created by a wire with current $I$ placed at the position of the replica, $(x,y) = (3/2 R_2,0)$}. Line plots of the normalized $y$-component (E) and $x$-component (F) of the magnetic induction along the dashed line in (A)-(D) ($x = 7/4 R_2$) for the cases $n_{\rm T} = 5$, $n_{\rm T} = 10$, $n_{\rm T} = 20$, and for the {replica} wire.  \label{Fig2}}
\end{figure}

For the experimental emulation of a current-carrying wire at a distance, we constructed a cylindrical {magnetic replicator} of external radius $R_2=40~$mm and height 400 mm. We set the {replica} wire current to $I=-0.5$A (the negative sign indicates the current flows in the negative $z-$direction) and its position to $d'=60$ mm. We truncated the summation in Eq. (1) up to $n_{\rm T}=10$. The resulting continuous sheet current density was converted into a discrete set of 20 straight current wires (current values and exact position of the wires can be found in the Supplementary Material). Figure 3A shows the field distribution obtained with numerical calculations when considering this discrete set of wires surrounding a centered wire of current $I=-0.5$ A. 

The magnetic field was measured in the rectangular region delimited in Fig. 3A using two miniature fluxgate magnetometers. The experimental results match very well with the results calculated numerically for the metamaterial device, which confirms that the magnetization currents of a shell with negative permeability can be used to create magnetic sources at a distance (Figs. 3B and 3C).

\begin{figure}[b!]
	\centering
	\includegraphics[width=0.5\textwidth]{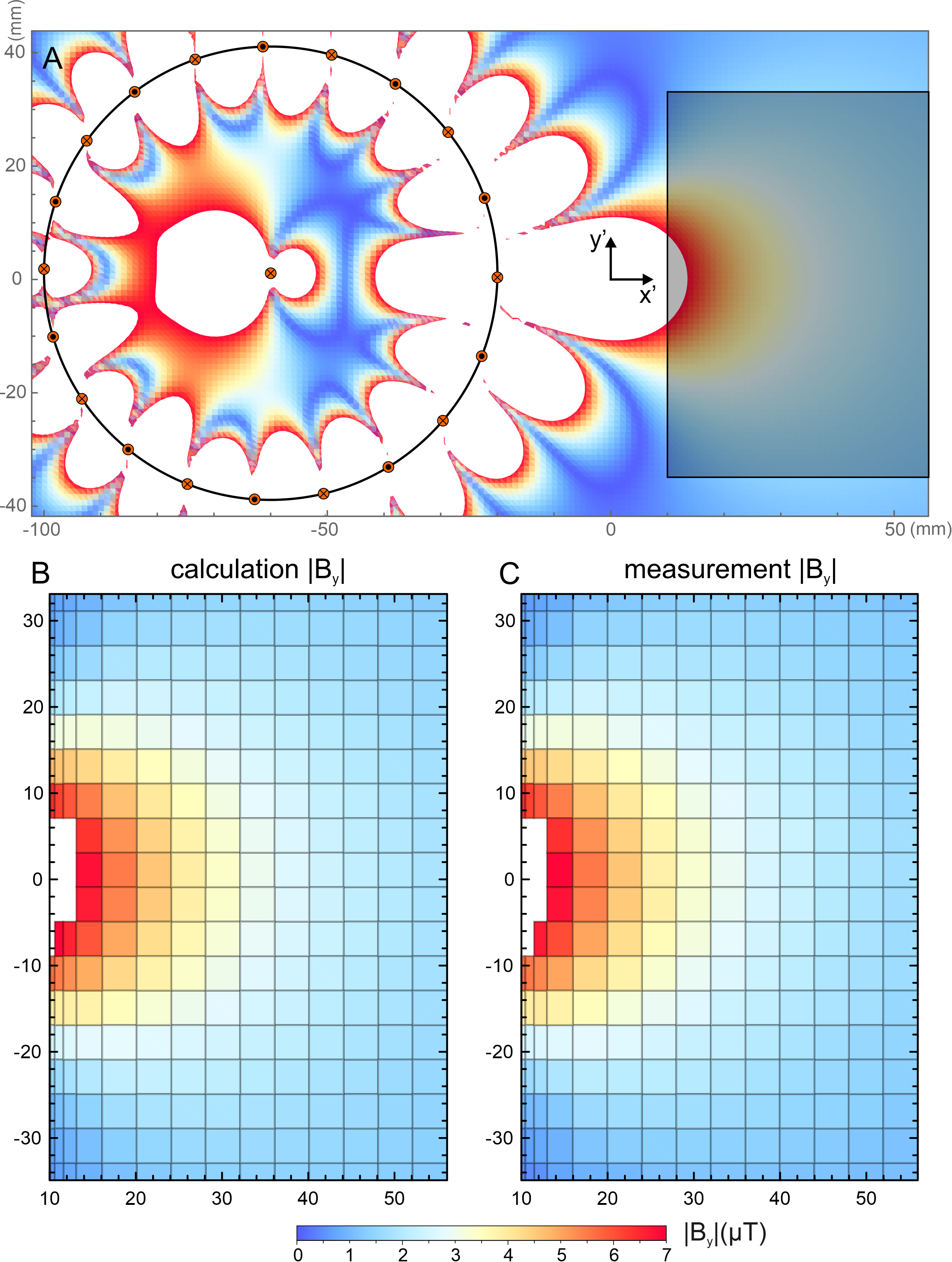} 
	\caption{Calculations and measurements of a {magnetic replicator}. (A) Plot of the calculated field $|B_y|$ created by the 21 currents (orange dots) forming the replicator with $R_2=40$mm and $d'=60$mm. The calculation assumes the wires have a length (in z) of 400mm. The shaded rectangle indicates the measured area. (B) $|B_y|$ field calculated at the positions of the measurements. (C) Measured field $|B_y|$. Standard errors associated with the measurements can be found in the Supplementary Material; values are smaller than $0.025\mu$T everywhere.\label{Fig3}}
\end{figure}
   
The {remote creation} of magnetic sources provides a new strategy for focusing magnetic fields in inaccessible volumes. Imagine a strong focusing of magnetic fields is required in a region where magnetic field sources cannot be placed, e.g. inside the body of a patient. One can place a {magnetic replicator} in the accessible volume (outside the body of that patient) in order to {emulate} the field of an image wire carrying a current $I$ close to the region where the focusing is required. Let us assume that the inaccessible region is $x>x_{\rm IN}$ and that the surface of the {magnetic replicator} is placed just at the border of the inaccessible region (the device is centered at the origin of coordinates and its radius is $R_2 = x_{\rm IN}$). The magnetic field distribution along different $y-$lines in the inaccessible region obtained using this strategy shows a much sharper peak than the field distribution that would be obtained by simply placing a wire with current $I$ at $x = x_{\rm IN}$ (Fig. 4A). The gradient of the field, $\partial B_y/\partial y$ is thus greatly increased. 
    
Moreover, we experimentally demonstrated that the field created by a magnetic source can be remotely cancelled by a {magnetic replicator}. Together with our metamaterial device, we placed a straight wire of current $I=0.5$ A in the position of the {replica wire (which we call the ``target wire")}. The superposition of the field generated by the wire and that generated by the metamaterial cancel out. The experimental measurements give a very low magnetic field strength in the region $\rho > d'$ (Fig. 4B). This demonstrates that magnetic sources placed in inaccessible regions (like the interior of a wall, for example) can be cancelled remotely, without the need for surrounding them with magnetic cloaks.

The {magnetic replicator} we propose, similar to parity-time perfect lenses for electromagnetic waves \cite{Fleury2014, Monticone2016}, does not require the design of bulk metamaterials with cumbersome magnetic permeability distributions; it can be realized by a precise arrangement of current-carrying electric wires. 
Thus, some of the issues associated with magnetic metamaterials, such as the non-linearity of their constituents~\cite{coey2010magnetism}, are avoided. 

Even though our results have been demonstrated for static magnetic fields, this same device could work for low-frequency AC fields \cite{Prat-Camps2016}. For time-dependent magnetic fields with associated wavelengths larger or comparable to the size of the device (for which the induced electric fields can be neglected), one could feed AC currents to the same active metamaterial. Thus, our experimental setup would maintain its functionality up to frequencies around 100MHz.

\begin{figure}[t!]
	\centering
	\includegraphics[width=0.5\textwidth]{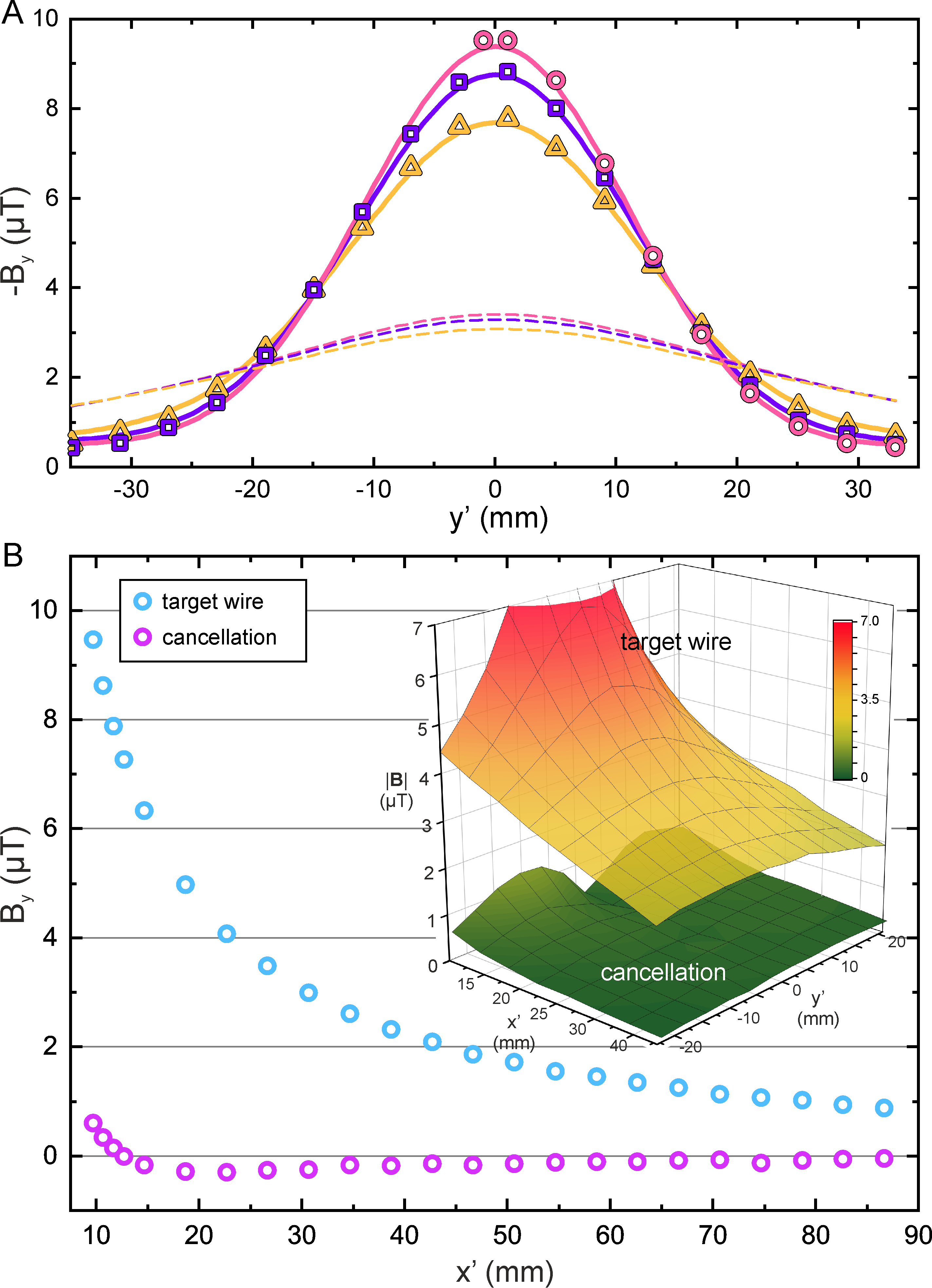} 
	\caption{Focusing of magnetic field and cancellation of the magnetic field created by a wire. (A) Plots of $-B_y$ along different parallel lines at {$x'=9.1, 10.1$ and $12.1$mm} (pink-circles, purple-squares, and yellow-triangles, respectively). 
		Measurements are shown in symbols and the corresponding calculations in solid lines. Dashed lines are calculations of the field created by a single finite wire carrying a current $I=-0.5$A located at $x'=-20$mm, $y'=0$. 
		(B) Measurements (line $y'=0$mm) of the field created by a wire located at the position of the image wire ("target wire") and of the field created by the {magnetic replicator} and the target wire ("cancellation"). Error bars (standard errors) are smaller than the symbol size. Insets show the measured $|\mathbf{B}|$ for the target wire and the cancellation configurations. The black grid shows the measurement points; the color surface is obtained as a linear interpolation between them. Standard errors are smaller than $0.02\mu$T. \label{Fig4}}
\end{figure}

The emulation of a magnetic source at distance is valid everywhere in space except in the circular region delimited by the position of the replica source and can be made as exact as desired at the expense of increased complexity and power requirements. 
According to Eq. (1), the higher the chosen number of terms $n_{\rm T}$, 
the larger the current density that has to be fed to the metamaterial shell. Since the summation in ${\bf K}^{n_{\rm T}}$ is dominated by its last term, the required number of wires to emulate the shell is $2 n_{\rm T}$. Not only its number but also the current each of them carries increases with $n_{\rm T}$; the current carried by the wire at $(x,y) = (R_2, 0)$, for example, increases with $n_{\rm T}$ as $\left(d'/R_2\right)^{n_{\rm T}}$. Therefore, the currents and the required power to create a source at a distance rapidly increase with the distance of the target source ($d'/R_2$) as well as with the total number of terms in ${\bf K}^{n_{\rm T}}$.
In scenarios where high field accuracy and strength are demanded, high-T$_c$ superconductors could be used to create the circuits forming the metamaterial.

In spite of considering translational symmetry along the $z-$direction throughout the article, these same ideas could be applied to emulate a 3D magnetic source, like a point magnetic dipole. In that case, one could consider a spherical shell with negative permeability and calculate the corresponding magnetization currents, which would likely result in cumbersome inhomogeneous distributions of surface and volume current densities.\\

Results presented here open a new pathway for controlling magnetic fields at a distance, with potential technological applications. 
For example, a wide variety of microrobots and functional micro/nano-particles are moved and actuated by means of magnetic fields~\cite{Sitti2009,Dreyfus2005,Kummer2010,Hu2018,Kim2018}. They can perform drug transport and controlled drug release~\cite{Estelrich2015}, intraocular retinal procedures~\cite{Kummer2010}, or even stem cell transplantation~\cite{Jeon2019}. However, the rapid drop off of field strength with target depth within the body is acknowledged to pose severe limitations to the clinical development of some of them~\cite{Estelrich2015,Sitti2009}. 
Another example is transcranial magnetic stimulation (TMS), which uses magnetic fields to modulate the neural activity of patients with different pathology~\cite{Wagner2007}. {In spite of its success, TMS suffers from limited focality, lacking the ability to stimulate specific regions}~\cite{Wagner2007}. 
Our results could benefit both technologies by enabling the precise spatial targeting of magnetic fields at the required depth inside the body. 
In actual applications, though, one should take into account that the region between the metamaterial and the {replica} image would experience strong magnetic fields.

Another area of application is in atom trapping. Depending on the atom's state, they can be trapped in magnetic field minima (low-field seekers) or maxima (high-field seekers). Since local maxima are forbidden by Earnshaw's theorem, high-field seekers are typically trapped in the saddle point of a magnetic potential that oscillates in time~\cite{Cornell1991}. However, these dynamic magnetic traps are shallow compared to traps for low-field seekers~\cite{Fortagh2007,Hinds_1999}. By emulating a magnetic source at distance, one would be able to generate magnetic potential landscapes with higher gradients at the desired target position resulting in tighter traps. 


In conclusion, our results demonstrate that a shell with negative permeability can emulate and cancel magnetic sources at distance. This ability to manipulate magnetic fields remotely will enable both the advancement of existing technology and potentially new applications using magnetic fields.

\section*{Acknowledgments}

JPC is funded by a Leverhulme Trust Early Career fellowship (ECF-2018-447). We thank P. Maurer, I. Hughes and C. Navau for their feedback on the article.


\newpage

\bibliographystyle{Science}
\bibliography{bibliography}

\newpage

\end{document}


\title{Supplementary Material for `Manipulating magnetic fields in inaccessible regions by negative magnetic permeability'}

\author{Rosa Mach-Batlle, Mark G. Bason, Nuria Del-Valle, Jordi Prat-Camps}

\date{}
\maketitle
\tableofcontents

\newpage

\section{A negative-permeability cylindrical shell by transformation optics}


\subsection{Space transformation}

\begin{figure}[h]
\centering
\includegraphics[scale=0.7]{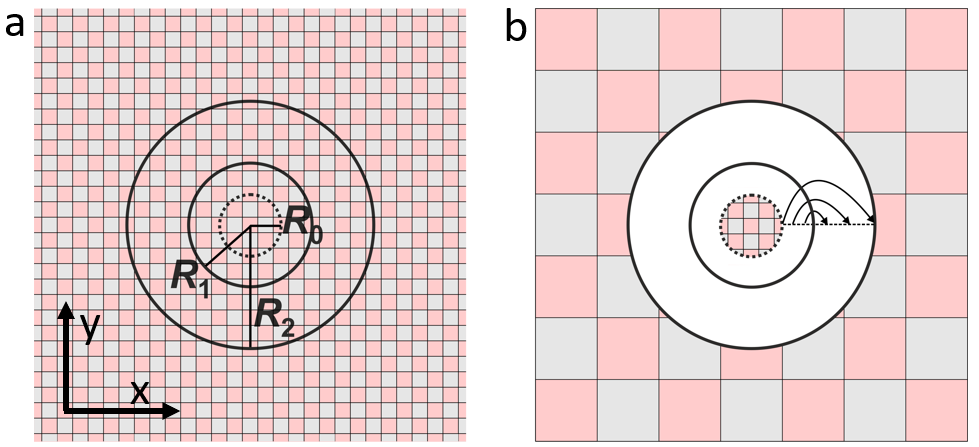} 
\caption{Sketches of (a) the undistorted original space and (b) the transformed space for magnetic field sources located inside the shell. \label{FigS2}}
\end{figure}

Our goal is to derive the properties of a cylindrical shell with negative permeability that acts on the magnetic field in the same way as a perfect lens acts on the electromagnetic waves (18, 31, 32). For internal magnetic fields, i. e. when a magnetic source is located inside the hole of the cylindrical shell, this shell must transform the space as illustrated in Fig. \ref{FigS2}.

Considering a cylindrical shell infinitely long along the $z-$direction with internal radius $R_1$ and external radius $R_2$ and a cylinder of radius $R_0 < R_1$,  the space is transformed as follows. The space in the region $R_0 < \rho < \infty$ is expanded through

\begin{equation}
\begin{cases}
\rho' = \left(\frac{R_2}{R_0}\right) \rho,\\
\varphi' = \varphi, \quad \quad \quad \quad  & \rho \in (R_0, \infty)\\
z' = z,\label{T1exp}
\end{cases}
\end{equation}
where $\rho, \varphi$ and $z$ are the coordinates in the original physical space and $\rho', \varphi'$ and $z'$ are the coordinates in the transformed virtual space. Simultaneously, the space $R_0 \leq \rho \leq R_1$ is folded (arrows in Fig. \ref{FigS2}b) as
\begin{equation}
\begin{cases}
\rho' = R_1 \left( \frac{\rho}{R_1}\right)^{k},\\
\varphi' = \varphi, \quad \quad \quad \quad  & \rho \in [R_0, R_1]\\
z' = z,\label{T2exp}
\end{cases}
\end{equation}
where $k$ is a negative parameter ranging from $0^-$ to $-\infty$. To guarantee the continuity of the space at $\rho' = R_2$, Eqs. \eqref{T1exp} and \eqref{T2exp} give the following relation between $k$ and $R_0$,
\begin{equation} \label{R0}
R_0 = R_1 (R_2/R_1)^{1/k}.
\end{equation}

Transformation optics theory can be applied to obtain the permeability resulting in the presented space transformation. The permeability tensors are
\begin{equation}
\mu' = 
\left( \begin{array}{ccc}
\mu_{\rho \rho} & \mu_{\rho \varphi} & \mu_{\rho z} \\
\mu_{\varphi \rho} & \mu_{\varphi \varphi} & \mu_{\varphi z} \\
\mu_{z \rho} & \mu_{z \varphi} & \mu_{z z} \end{array} \right)
=
\left( \begin{array}{ccc}
1 & 0 & 0 \\
0 & 1 & 0 \\
0 & 0 & \left(\frac{R_2}{R_1} \right)^{2/k-2} \end{array} \right)\label{TOair}  
\end{equation}
in the external region $\rho'>R_2$,
\begin{equation}
\mu' = \left( \begin{array}{ccc} 
k & 0 & 0 \\
0 & 1/k & 0 \\
0 & 0 & \frac{1}{k}\left(\frac{\rho'}{R_1}\right)^{2/k -2} \end{array} \right) \label{TOmaterial}
\end{equation}
in the cylindrical shell region $R_1 \leq \rho' \leq R_2$, and $\mu' = 1$ in the internal region, $\rho'<R_1$, where the space is not transformed. 

Because we are considering static magnetic fields and in magnetostatics the magnetic and the electric fields are decoupled, the permittivity $\varepsilon$ of the medium does not play any role in the shaping of magnetic fields. 
%
%
If one assumes translational symmetry along the $z-$axis, the $z-$component of the magnetic field is zero and only the left-upper 2x2 minors of the permeability tensors have physical relevance. In this case, Eq. \eqref{TOair} shows that a negative-permeability shell, different from the perfect lenses proposals for electromagnetic waves (18,31,32), does not require magnetic material neither inside the hole nor in the external region. According to Eq. \eqref{TOmaterial}, the relative magnetic permeability of a negative-permeability cylindrical shell is homogeneous and anisotropic, with radial and angular components $\mu_{\rho \rho} = k$ and $\mu_{\varphi \varphi} = 1/k$, respectively. Since $k$ was defined as a negative parameter, there are infinite possible anisotropic shells, all with negative values of  both $\mu_{\rho \rho}$ and $\mu_{\varphi \varphi}$. All these cylindrical shells exhibit slightly different properties, since they result from different space transformations [Eqs. \eqref{T1exp}-\eqref{R0}]. 


Similar space transformations demonstrate that this anisotropic long cylindrical shell with  negative radial and angular components of the relative magnetic permeability $k$ and $1/k$, respectively,  transforms externally applied magnetic fields in an analogous way \cite{Navau2012}.


\subsection{Analytic expressions of the magnetic field}

One can apply the transformation optics theory to find out how a negative-permeability shell transforms a magnetic field distribution everywhere in space.
%
%
The analytic expression of the magnetic field when the space is transformed according to Eqs. \eqref{T1exp} and \eqref{T2exp} is
\begin{equation}
{\bf H}'(\rho',\varphi') = {\bf H}(\rho', \varphi'), \quad \quad \quad \quad \quad \quad \quad \quad \quad \quad \quad \quad  \quad \quad  \quad \, \, \rho' \in [0,R_1) \label{TOHINT}
\end{equation}
\begin{equation}
\begin{cases}
{H}_\rho'(\rho',\varphi') = \frac{1}{k} \left(\frac{\rho'}{R_1} \right)^{1/k-1} H_\rho \left(R_1 \left(\frac{\rho'}{R_1} \right)^{1/k}, \varphi' \right), \\
{H}_\varphi'(\rho',\varphi') = \left(\frac{\rho'}{R_1} \right)^{1/k-1} H_\varphi \left(R_1 \left(\frac{\rho'}{R_1} \right)^{1/k}, \varphi' \right), \quad \quad \quad  \quad \rho' \in [R_1,R_2] 
\end{cases}\label{TOHSHE}
\end{equation}
\begin{equation}
{\bf H}'(\rho',\varphi') = \left(\frac{R_2}{R_1} \right)^{1/k-1} {\bf H} \left( \left(\frac{R_2}{R_1} \right)^{1/k-1} \rho', \varphi' \right).  \quad \quad  \quad \quad   \rho' \in (R_2, \infty) \label{TOHEXT}
\end{equation}
Equation \eqref{TOHINT} shows that the field inside the hole, $\rho' \in [0,R_1)$, is not modified by the presence of the shell, while Eqs. \eqref{TOHSHE} and \eqref{TOHEXT} indicate that the field distribution in the shell volume, $\rho' \in [R_1,R_2]$, and in the external region, $\rho' \in (R_2, \infty)$, are transformed by the shell and depend on the magnetic field source and on the shell radii and parameter $k$.

\subsection{Transformation of the magnetic field created by an inner wire} 

\begin{figure}[h]
\centering
\includegraphics[scale=0.6]{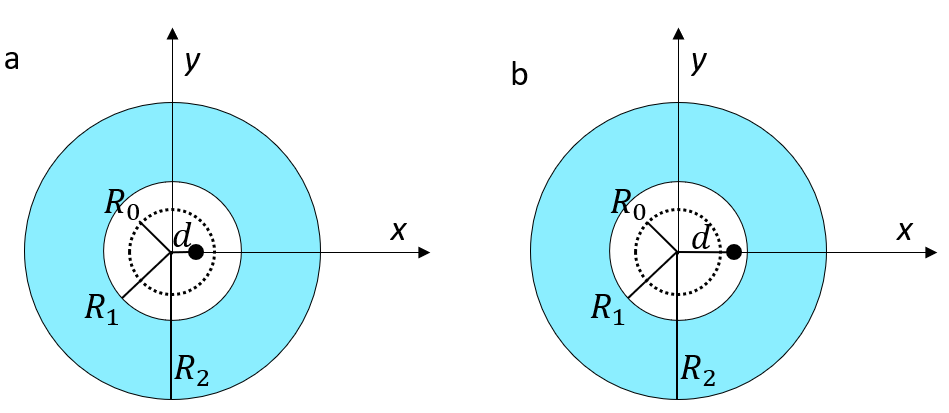} 
\caption{Sketches of two long (along $z$) cylindrical negative-permeability shells (blue regions) with internal radius $R_1$ and external radius $R_2$ surrounding a long straight wire of current $I$ (black dot) placed at a distance $d$ from the center of the shell. According to the space transformation in Eqs. \eqref{T1exp} and \eqref{T2exp}, the shell effectively folds the space in the region $\rho \in [R_0, R_1]$. The radius $R_0$ is plotted as a dotted line. In (a) the wire is located in the non-folding region ($d < R_0$), while in (b) it is located in the folding region ($R_0 <d < R_1$). \label{FigS3}}
\end{figure}

Here we use the general expressions in Eqs. \eqref{TOHINT}-\eqref{TOHEXT} to derive how a negative-permeability shell transforms the magnetic field generated by a long straight wire of dc current $I$ placed inside its hole, at a distance $d$ from the center of the shell (see Fig. \ref{FigS3}). Other magnetic field distributions with translational symmetry along the $z-$direction, such as the field created by a long magnetic dipole or a long magnetic quadrupole, can be regarded as the superposition of the field created by several wires. For example, the field created by two antiparallel wires of current $I$ separated by a small distance $d \to 0$ approximates well the field created by a dipolar source with dipolar magnetic moment per unit length $m = I d$. Therefore, the transformation of the field created by a wire can be used to understand how negative-permeability shells transform these other field distributions.

Because the shell does not transform the field inside its hole, the field in the region $\rho < R_1$ corresponds to the field of the original wire. In the exterior of the shell, the field distribution is found to be equivalent to the field of a replica wire of current $I$ located at
\begin{equation}\label{rho'}
d^{\rm EXT}_{\rm i} = \left(\frac{R_2}{R_1}\right)^{1-1/k} d.
\end{equation}
If the original wire is placed outside the folding region ($d < R_0$, as in Fig. \ref{FigS3}a), the replica wire appears in the shell region $d^{\rm EXT}_{\rm i} < R_2$. In contrast, if the original wire is placed in the folding region ($R_0 < d < R_1$, as in Fig. \ref{FigS3}b), the replica wire appears at a distance from the external surface of the shell, at $d^{\rm EXT}_{\rm i} > R_2$. However, the magnetostatic Maxwell equations $\nabla \times {\bf H} = {\bf J}_{\rm f}$ and $\nabla \cdot {\bf B} = 0$ indicate that magnetic field lines can only emanate from actual magnetic sources. This makes it impossible to achieve the same field distribution as if a source was located in empty space (at $d^{\rm EXT}_{\rm i} > R_2$) if no real source is actually placed at $d^{\rm EXT}_{\rm i}$. Therefore, the transformation optics results for the cases in which the folding region contains magnetic sources require the consideration of actual sources at the replica location. Perfect lenses for electromagnetic waves derived by transformation optics show similar limitations \cite{Bergamin2010}.

In the shell region, the magnetic field can only be written in terms of replica wires when the permeability of the shell is isotropic, with $\mu_{\rho \rho} =  \mu_{\varphi \varphi} = -1$. In this case, the magnetic field ${\bf H}$ in the shell region corresponds to the superposition of the field of a replica wire of current $I$ located at the center of the shell hole plus another one of current $-I$ placed at $d^{\rm SHE}_{\rm i}=R_1^2/d$. Because $\mu = -1$ and ${\bf B} = \mu\mu_0{\bf H}$, the magnetic induction ${\bf B}$ in the shell is that of a replica wire of current $-I$ located $\rho = 0$ plus that of another one of current $I$ placed at $d^{\rm SHE}_{\rm i}$.

\newpage
\section{A negative-permeability cylindrical shell by Maxwell equations}

\subsection{Analytic derivation of the magnetic vector potential for a negative-permeability shell surrounding a wire}

Consider a cylindrical shell of internal radius $R_1$ and external radius $R_2$ with linear, homogeneous and isotropic relative magnetic permeability $\mu = -1$, which we coin a \textit{magnetic replicator}. It surrounds a long straight wire of current $I$ displaced a distance $d$ from the center of the hole, as illustrated in Fig. \ref{FigS3}.

The magnetic field everywhere in space can be derived from the magnetic vector potential ${\bf A}$ as ${\bf B} = \nabla \times {\bf A}$. Due to the symmetry of the field created by the wire, the magnetic vector potential is found to be ${\bf A} = A(\rho, \varphi) {\bf z}$. From the magnetostatic Maxwell equation $\nabla \times {\bf H} = {\bf J}_{\rm f}$ and the constitutive relation ${\bf B} = \mu \mu_0 {\bf H}$, one finds that (in the Coulomb Gauge; i.e. $\nabla \cdot {\bf A} = 0$) the magnetic vector potential must fulfill
\begin{equation}\label{EqA}
\nabla^2 A(\rho, \varphi) = 0,
\end{equation}
inside the hole (INT), in the magnetic replicator volume (MR), and in the exterior of the shell (EXT) while
\begin{equation}
\nabla^2 A(\rho, \varphi) = -\mu_0 J_{\rm f},
\end{equation}
in the region occupied by the wire, which we consider to be a point.

The solution of Eq. \eqref{EqA} can be written as,
\begin{align}
A^{\rm INT}(\rho, \varphi) = A_{\rm w}(\rho, \varphi) + \sum_{n =1}^{\infty} a_n \rho^n {\rm cos}(n \varphi)\\ 
A^{\rm MR}(\rho, \varphi) = b_0{\rm ln}(\rho) + \sum_{n =1}^{\infty} \left(b_n \rho^n + \frac{c_n}{\rho^n} \right) {\rm cos}(n \varphi), \label{EqAPL}\\
A^{\rm EXT}(\rho, \varphi) = d_0{\rm ln}(\rho) + \sum_{n =1}^{\infty}  \frac{d_n}{\rho^n} {\rm cos}(n \varphi),
\end{align}
where $A_{\rm w}$ is the magnetic vector potential due to a non-centered long straight wire and we have taken into account that the magnetic field the shell creates must be finite at $\rho \to 0$ and tend to zero at $\rho \to \infty$. In order to obtain the magnetic vector potential coefficients it is convenient to write the magnetic vector potential of a wire $A_{\rm w}$ as a Fourier series, which reads as
\begin{equation}\label{phiD}
A_{\rm w}(\rho, \varphi) = 
\begin{cases} 
-\frac{\mu_0 I}{2 \pi}{\rm ln}(\rho) + \sum_{n=1}^{\infty} \frac{\mu_0 I}{2 \pi n} \left(\frac{d}{\rho}\right)^n {\rm cos}(n\varphi), \quad \quad & \rho > d \\\\
-\frac{\mu_0 I}{2 \pi}{\rm ln}(d) + \sum_{n=1}^{\infty} \frac{\mu_0 I}{2 \pi n} \left(\frac{\rho}{d}\right)^n {\rm cos}(n\varphi). \quad \quad & \rho \leq d \\
\end{cases}
\end{equation}

The coefficients $b_0$, $d_0$, $a_n$, $b_n$, $c_n$, and $d_n$, obtained by imposing that the radial component of the magnetic induction and the angular component of the magnetic field must be continuous at the material boundaries $\rho = R_1$ and $\rho = R_2$, are
\begin{align}
b_0 &= \frac{\mu_0 I}{2 \pi},\\
d_0 &= -\frac{\mu_0 I}{2 \pi},\\
a_n &= 0,\\
b_n &= \frac{\mu_0 I}{2\pi n} \left(\frac{d}{R_1^2}\right)^n,\label{Eqbn}\\
c_n &= 0,\\
d_n &= \frac{\mu_0 I}{2\pi n} \left(\frac{d R_2^2}{R_1^2} \right)^n.
\end{align}

\begin{figure}[h]
\centering
\includegraphics[width=1\textwidth]{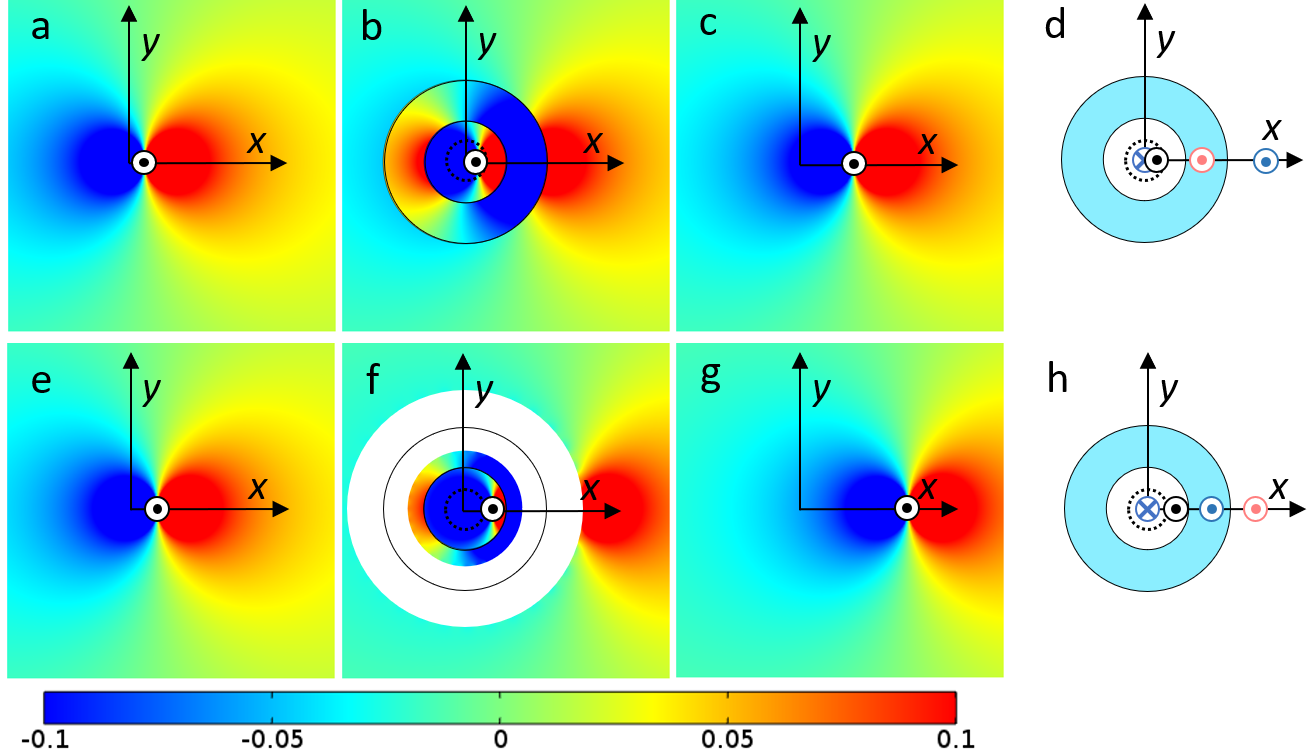} 
\caption{In panels (a)-(c) and (e)-(g), numerical calculations of the normalized magnetic induction $B_y R_1/(\mu_0 I)$ a straight wire of current $I$ (sketched as a black dot) creates in different configurations. In (a) the wire is placed at ($x$,$y$)= ($d$, 0). In (b) the wire in (a) is surrounded by a cylindrical magnetic replicator with radii $R_1 = 3d$ and $R_2 = 2R_1$ and relative magnetic permeability $\mu = -1$ centered at the origin of coordinates. In (c) the wire is located at ($4d$, 0). In (e) the wire is placed at ($2d$, 0). In (f) the wire in (e) is surrounded by the same shell as in (b). In (g) the wire is placed at ($8d$, 0). The white annular regions in (f) indicate that the field diverges. Panels (d) and (h) illustrate the creation of replicas achieved by the magnetic replicator in (b) and (f), respectively. The replica wire sketched in red in (d) and (h) gives the field distribution in the external region, while the replica wires sketched in blue give the field distribution in the shell volume. 
 \label{figwire}}
\end{figure}

Some general properties of the magnetic replicator result from the magnetic vector potential solutions. First, because $a_n = 0$, the shell does not distort the magnetic field created by the wire inside its hole (compare Figs. \ref{figwire}b and f to Figs. \ref{figwire}a and e, respectively). Second, when $d \leq R_1^2/R_2$, the magnetic vector potential in the external region converges to the potential of a replica current $I$ located at $d^{\rm EXT}_{\rm i} = d(R_2/R_1)^2$ (compare Figs. \ref{figwire}b and c). If $d > R_1^2/R_2$, the replica current appears outside the shell ($d^{\rm EXT}_{\rm i} > R_2$) and $A^{\rm EXT}(\rho, \varphi)$ converges to the potential of the replica wire not in the whole external region but only in the region $\rho > d^{\rm EXT}_{\rm i}$; in the region $\rho \in (R_2 , d^{\rm EXT}_{\rm i})$ $A^{\rm EXT}(\rho, \varphi)$ diverges (compare Figs. \ref{figwire}f and g). Third, when $d \leq R_1^2/R_2$, the  magnetic vector potential in the shell region converges to the potential of a replica wire of current $-I$ placed at $\rho = 0$ plus the potential of a second replica wire of current $I$ placed at $d^{\rm SHE}_{\rm i} = R_1^2/d$ (outside the shell, $d^{\rm SHE}_{\rm i} > R_2$). If $d > R_1^2/R_2$, the second image appears in the shell volume ($R_1 < d^{\rm SHE}_{\rm i} < R_2$). In this case, $A^{\rm MR}(\rho, \varphi)$ converges to the potential of the two replica wires in the region $\rho \in (R_1, d^{\rm SHE}_{\rm i})$ and diverges in the region $\rho \in (d^{\rm SHE}_{\rm i}, R_2)$. 

\subsection{Mimicking the magnetic replicator with current distributions}

The magnetic replicator requires a negative value of the magnetic permeability. Materials with negative permeabilities do not naturally occur, but their behaviour can be effectively mimicked by substituting the hypothetical material with negative permeability by its magnetization currents (12). Here we derive the magnetization currents of a cylindrical magnetic replicator with permeability $\mu = -1$ surrounding a long straight wire, as sketched in Fig. \ref{FigS3}. Because the material is homogeneous and isotropic, the volume magnetization currents are zero, ${\bf J}_{\rm M}(\rho,\varphi) = \nabla \times {\bf M}(\rho,\varphi) = 0$. The surface magnetization current densities ${\bf K}_{\rm M_1}(\varphi)$ and ${\bf K}_{\rm M_2}(\varphi)$, flowing at $\rho = R_1$ and $\rho = R_2$ respectively, are found from the material magnetization as,

\begin{equation}
{\bf K}_{\rm M_1}(\varphi) = {\bf M}(R_1, \varphi) \times {\bf n}_1 = M_\varphi(R_1, \varphi){\bf z} = \frac{2}{\mu_0}B_\varphi(R_1, \varphi){\bf z} = -\frac{2}{\mu_0}\frac{\partial A^{\rm MR}(\rho, \varphi)}{\partial \rho}\biggr\rvert_{\rho=R_1}{\bf z},
\end{equation}\\

\begin{equation}
{\bf K}_{\rm M_2}(\varphi) = {\bf M}(R_2, \varphi) \times {\bf n}_2 = -M_\varphi(R_2, \varphi){\bf z} = -\frac{2}{\mu_0}B_\varphi(R_1, \varphi){\bf z} = \frac{2}{\mu_0}\frac{\partial A^{\rm MR}(\rho, \varphi)}{\partial \rho}\biggr\rvert_{\rho=R_2}{\bf z},
\end{equation}
where ${\bf n}_1$ and ${\bf n}_2$ are the unity vectors perpendicular to the surfaces $\rho =R_1$ and $\rho =R_2$, respectively, pointing outwards the material surfaces; ${\bf n}_1 = -{\pmb{\rho}}$ and ${\bf n}_2 = {\pmb{\rho}}$. From Eqs. \eqref{EqAPL} and \eqref{Eqbn}, the surface magnetization current densities are obtained as 

\begin{equation}
{\bf K}_{\rm M_1}(\varphi) = \left(-\frac{I}{\pi R_1} - \sum_{n=1}^{\infty} \frac{I}{\pi R_1}\left(\frac{d}{R_1}\right)^n{\rm cos}(n \varphi) \right) {\bf z},
\end{equation}

\begin{equation}
{\bf K}_{\rm M_2}(\varphi) = \left( \frac{I}{\pi R_2} + \sum_{n=1}^{\infty} \frac{I}{\pi R_2}\left(\frac{d R_2}{R_1^2}\right)^n{\rm cos}(n \varphi) \right) {\bf z}.
\label{eq31} \end{equation}

The sum in the current density ${\bf K}_{\rm M_1}$ is always convergent and gives

\begin{equation}
{\bf K}_{\rm M_1}(\varphi) = \left(-\frac{I}{\pi R_1} + \frac{I}{\pi R_1}\frac{d[d-R_1{\rm cos}(\varphi)]}{R_1^2+d^2-2dR_1{\rm cos}(\varphi)} \right) {\bf z}.\label{eq32}
\end{equation}

In contrast, the current density ${\bf K}_{\rm M_2}$ only converges when $d< R_1^2/R_2$. In this case,

\begin{equation}
{\bf K}_{\rm M_2}(\varphi) = \left( \frac{I}{\pi R_2} - \frac{I}{\pi}\frac{d[d R_2-R_1^2{\rm cos}(\varphi)]}{R_1^4+d^2R_2^2-2dR_1^2R_2{\rm cos}(\varphi)} \right)  {\bf z}.
\end{equation}

When the summation in Eq. \eqref{eq31} is not convergent one can still approximately obtain the same field distribution as that resulting from a magnetic replicator with permeability $\mu = -1$ by truncating the external current density summation up to $n_{\rm T}$ terms,

\begin{equation}
{\bf K}^{\rm n_T}_{\rm M_2}(\varphi)= \left(\frac{I}{\pi R_2} + \sum_{n=1}^{n_{\rm T}} \frac{I}{\pi R_2}\left(\frac{d R_2}{R_1^2}\right)^n{\rm cos}(n \varphi) \right) {\bf z}.\label{eq34} \end{equation}

As shown in Fig. 2 of the main text, the higher the number of terms ${n_{\rm T}}$, the more the external field distribution resembles the field of the replica wire. 

We can now analyse what is the field distribution that the current densities ${\bf K}_{\rm M_1}(\varphi)$ and ${\bf K}_{\rm M_2}(\varphi)$ create in the external region. On the one hand, ${\bf K}_{\rm M_1}(\varphi)$ cancels the field created by the original wire and creates the same field distribution as that of a centered wire of current $-I$. On the other hand, ${\bf K}_{\rm M_2}(\varphi)$ cancels the field created by a centered wire of current $-I$ and creates the field of the replica wire. In this way when the original wire is surrounded by the two current densities ${\bf K}_{\rm M_1}(\varphi)$ and  ${\bf K}_{\rm M_2}(\varphi)$ the field in the external region corresponds to the field created by a replica wire. 

\begin{figure}[h]
\centering
\includegraphics[width=1\textwidth]{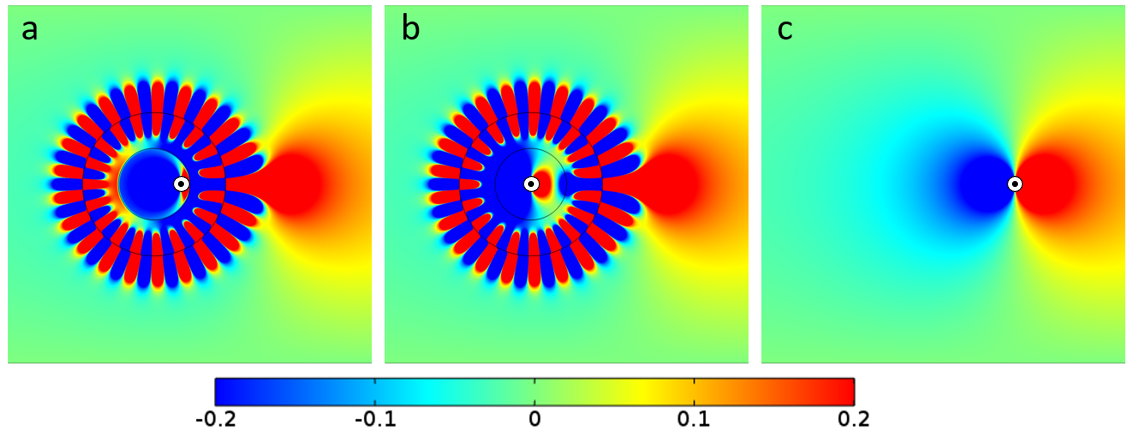} 
\caption{Normalized vertical component of the magnetic induction $B_y R_2/(\mu_0 I)$ for (a) a wire of current $I$ placed at $x=d$ surrounded by the sheet current densities in Eqs. \eqref{eq32} and \eqref{eq34} with $R_1 = 4/3d$, $R_2 = 2R_1$, and $n_{\rm T} = 20$, (b) a centered wire of current $I$ surrounded by the current density in Eq. \eqref{eq36} with $R_2 = 8/3d$, $d_{\rm i}^{\rm EXT} = 4d$ and $n_{\rm T} = 20$, and (c) a wire of current $I$ placed at $x=4d$. \label{fig4sm}}
\end{figure}

If the goal is to create the field of a wire in the external region regardless of the field in the volume occupied by the shell, the original wire and the current density ${\bf K}_{\rm M_1}(\varphi)$ can be substituted by a centered wire of current $-I$; this centered wire together with ${\bf K}_{\rm M_2}(\varphi)$ still create the field of a replica wire. Actually, as shown in Fig. \ref{fig4sm}, the currents can be simplified even further. Since the first term in the summation of ${\bf K}_{\rm M_2}(\varphi)$ creates the same field distribution in the external region as a centered wire of current $2I$, the field of the replica wire can be obtained by a centered wire of current $I$ surrounded by the surface current density

\begin{equation}
{\bf K}^{\rm n_T}(\varphi) = \sum_{n=1}^{n_{\rm T}} \frac{I}{\pi R_2}\left(\frac{d R_2}{R_1^2}\right)^n{\rm cos}(n \varphi)  {\bf z},\label{eq35} \end{equation}
which can be written in terms of the position of the replica wire as 

\begin{equation}
{\bf K}^{\rm n_T}(\varphi) = \sum_{n=1}^{n_{\rm T}} \frac{I}{\pi R_2}\left(\frac{d_{\rm i}^{\rm EXT}}{R_2}\right)^n{\rm cos}(n \varphi)  {\bf z}.\label{eq36} \end{equation}

\subsection{Design of an active metasurface that emulates a wire at a distance}

Consider a magnetic replicator that is designed to create a target source at a distance, i.e. the chosen parameters fulfill $R_1^2/R_2<d<R_1<R_2<d_{\rm i}^{\rm EXT}$. For a practical realization of such a device, the continuous surface current density in Eq. \eqref{eq36} must be converted into a discrete set of straight wires constituting an active metamaterial or metasurface. The current of each of the wires can be calculated according to
\begin{equation}\label{I}
    I_{\rm m}=\int_{(\theta_{m}+\theta_{m-1})/2}^{(\theta_{m+1}+\theta_{m})/2} {K}^{n_{\rm T}}(\varphi) R_2 d\varphi,
\end{equation}
where $\theta_m$ is the angular positions of each wire. 

If one chooses to place a wire at the positions where the function ${\bf K}^{n_{\rm T}}(\varphi)$ has a relative minimum or a maximum, the metasurface must consists of 2$n_{\rm T}$ straight wires, as the summation in Eq. \eqref{eq36} is dominated by its last term. Therefore, the creation of a replica wire of current $I$ requires 2$n_{\rm T}$ straight wires of current $I_{\rm m}$ ($ m = 1,2,3...2n_{\rm T}$) plus a centered wire of current $I_{2n_{\rm T}+1} = I$.

According to Eq. \eqref{eq36}, the higher the chosen number of terms $n_{\rm T}$, the larger the current density. The number of straight wires and the current $I_m$ of each of the wires also increases with $n_{\rm T}$. For example, the current of the wire that is placed at $(x,y) = (R_2, 0)$, which is the wire that is placed closer to the replica source and the one that must carry the largest current, increases with $n_{\rm T}$ as $\left(d_{\rm i}^{\rm EXT}/R_2\right)^{n_{\rm T}}$, approximately. Because the currents required to create a source at a distance increase with $n_{\rm T}$, the power dissipated by the metasurface, which can be written as the summation of the power dissipated by each of the wires,
\begin{equation}
    P_{\rm T} = \sum_{m=1}^{2n_{\rm T}+1} I_m^2R_m,
\end{equation}
also increases with $n_{\rm T}$. 

We can now compare the power dissipated by the metamaterial magnetic replicator that creates the field of a replica wire at a distance, $P_{\rm T}$, for different $n_{\rm T}$ to the power that would be dissipated by an actual replica wire, $P_0$. Assume a long straight wire of current $I_{2n_{\rm T}+1} = I$ placed at the origin of coordinates surrounded by a cylindrical metasurface of radius $R_2$ consisting of different sets of long straight wires of current $I_{\rm m}$. The currents $I_{\rm m}$ are calculated from Eqs. \eqref{eq36} and \eqref{I} assuming $d_{\rm i}^{\rm EXT}=1.5R_2$ and 5, 10 and 20 $n_{\rm T}$ terms. 

Consider first that we have 2$n_{\rm T}$ long straight wires connected to the same voltage source, which provides a constant voltage $V_{\rm m} = V$. We assume that the resistance of each wire can be tuned so that each of them carries the appropriate current, $I_m$. The total power dissipated by the central wire and the $2n_{\rm T}$ wires in the cylindrical surface normalized to the power that would be dissipated by the replica wire ($P_0 = I V$) is
\begin{equation}
\frac{P_{\rm T}}{P_0} = \frac{1}{I V}\sum_{\rm m=1}^{\rm 2n_T+1} I_m V_m = \sum_{\rm m=1}^{\rm 2n_T+1}  \frac{I_m }{ I }. \end{equation}
The values of $P_{\rm T}/P_0$ obtained for $n_{\rm T}=5$, $n_{\rm T}=10$, and $n_{\rm T}=20$ are 11.13, 84.59 and 4877.69, respectively. The higher the number of terms that are considered (and therefore, the more the field resembles that of a source appearing at a distance), the larger the power needed to feed the metamaterial. 

Consider now that we followed a different strategy for feeding the current to all the wires. Imagine we had a single wire of current $I_{\rm c}$ and that we used this wire to achieve all the currents $I_{\rm m}$ in the experiment by adjusting the number of turns at each angular position. In this way, the dissipated power normalized to the power dissipated by the replica wire ($P_0 = I V = I^2 R$) would be calculated as
\begin{equation}
\frac{P_{\rm T}}{P_0} = \frac{1}{I^2 R}\sum_{\rm m=1}^{\rm 2n_T+1} I^2_{\rm m} R_{\rm m} = \sum_{\rm m=1}^{\rm 2n_T+1} \frac{I_{\rm m}^3}{I^3},
\end{equation}
taking into account that the resistance $R_{\rm m}$ is proportional to the number of turns of the wire of current $I_{\rm c}$ required for achieving $I_{\rm m}$. For the cases $n_{\rm T}=5$, $n_{\rm T}=10$, and $n_{\rm T}=20$, the obtained values of $P_{\rm T}/P_0$ are 45.73, 4531.72, and $1.93\cdot10^8$, respectively, which show the strong dependence of the power dissipated on the number of terms $n_{\rm T}$ that is considered and also on the chosen experimental method for feeding the currents.

\newpage
\section{Materials and Methods}

\subsection{Design and construction of the magnetic replicator} \label{sec.experimental}

\begin{figure}[b!]
\centering
\includegraphics[width=1.0\textwidth]{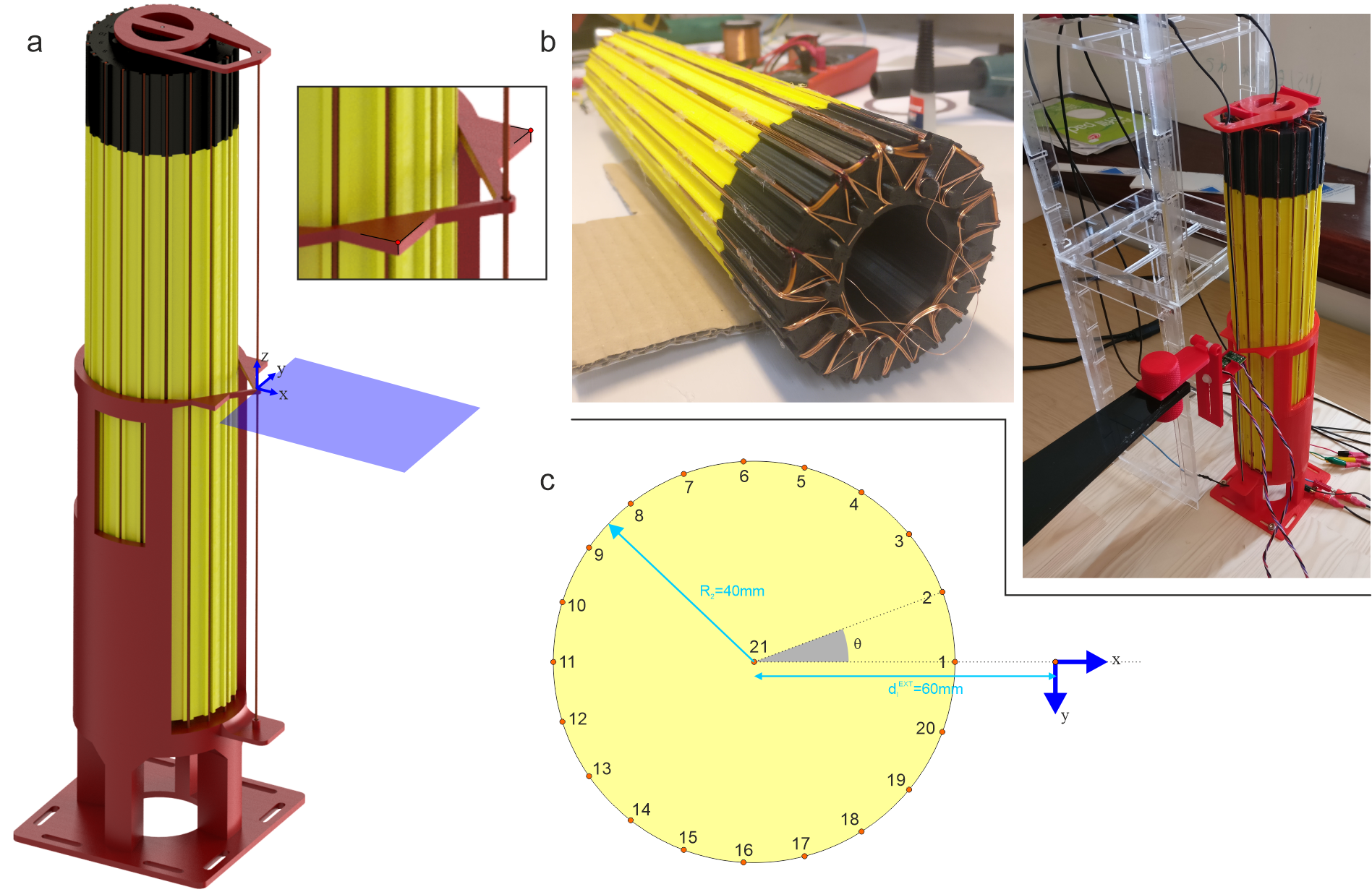} 
\caption{(a) 3D render of the experimental setup. Red, yellow, and black parts represent different 3D-printed support structures to position the wires (shown in orange-copper color) at the desired locations. The scan area is shown in translucent blue. The inset shows the two vertices that were used to determine the relative position of the magnetic replicator with respect to the sensors. (b) Pictures of the experimental device, showing the connections between different turns (left) and the actual experimental setup. (c) Sketch indicating the position of the different wires and the definition of the angle $\theta$.  \label{Figrender}}
\end{figure}

We designed and built a cylindrical magnetic replicator that emulates a wire at a certain distance from the shell. In particular, our replicating shell has an external radius $R_2=40$mm and generates a replica current at $d_{\rm i}^{\rm EXT}=60$mm (see Fig. \ref{Figrender}c).
Since the replica wire appears outside the shell, the required ${\bf K}_{\rm M_2}$ would not converge. For this reason, we truncate the summation in Eq.~\eqref{eq36} up to $n_{\rm T}=10$. The resulting continuous sheet current density is converted into a discrete set of 20 straight current wires. The angular position of these wires is chosen to be at the positions where the function ${\bf K}^{n_{\rm T}}(\varphi)$ has a relative minimum or a maximum. The current each of these wires carries is given by
\begin{equation}\label{I}
    I_{m}=\int_{(\theta_{m}+\theta_{m-1})/2}^{(\theta_{m+1}+\theta_{m})/2} {K}^{n_{\rm T}}(\varphi) R_2 d\varphi,
\end{equation}
where $\theta_m$ are the angular positions of each wire given in Table~\ref{table1}. The current of each wire ($I_m$) is also given. These values are obtained assuming the replica wire carries a current $I=-0.5$A (the negative sign indicates the current flows in the negative z-direction). 
The sheet current density ${\bf K}_{\rm M_1}$ converges to a finite number. As discussed in the previous section, if we are interested in the field distribution outside the shell only, this current density can be replaced by a single wire placed at the center of the cylindrical replicator. This central wire is referred as $m=21$ in Table~\ref{table1}.

Table~\ref{table1} indicates that 12 different values of current are needed. We used only 4 independent circuits, each of them fed with a different value of current, to generate all the values. We combined different number of turns of each circuit  to produce the required effective current as described in Table~\ref{table1} (negative values indicate turns in the negative direction of the z axis). Circuits 1, 2 and 3 were fed with $2.5$, $0.25$, and $0.025$A, respectively, and were combined to generate the current for the 20 external wires. Circuit 4 carried a current of 0.5A and fed the central wire ($m$=21) only. Circuits 1, 2, and 3 consisted of single enameled copper wires of 1.59, 0.50 and 0.22mm of diameter, respectively. Circuit 4 used a single AWG wire of 1.5mm of diameter. 

The actual experimental magnetic replicator was designed to have a length (along the z direction) of 400mm. In order the position all the wires at the appropriate positions, a non-magnetic plastic structure was designed and 3D-printed (using a FlashForge Creator Pro printer and polylactic acid [PLA] filament). This cylindrical structure (shown in yellow and black in Fig.~\ref{Figrender}a-c) had grooves at the appropriate angular positions, where the different turns where positioned and glued. Each of the 4 circuits was winded continuously, as can be seen in Fig.~\ref{Figrender}b. The central wire was guided through the central hole of the cylinder using specially-designed 3D-printed parts placed at the top and the bottom of the cylinder. In a similar way, a wire was placed at the position of the replica wire. This "target" wire was only fed in some parts of the experiment to demonstrate, for example, the cancellation properties of our magnetic replicator.
Finally, a 3D-printed support structure (shown in red in Fig.~\ref{Figrender}a and c) held the structure upright and clamped it to the table. Two reference points of this structure (see inset in Fig.~\ref{Figrender}a) were used to calibrate the relative position of the whole structure with respect to the magnetic sensors. 
For this point on, the origin of coordinates is set at the position of the target wire as depicted in Fig.~\ref{Figrender}a. All the measurements and simulations shown next follow this convention.


\begin{table}[h!]
\centering
\small
\begin{tabular}{c|c|c|c|c|c|c}
\begin{tabular}[c]{@{}c@{}} wire number, \\ m\end{tabular}  & \begin{tabular}[c]{@{}c@{}}$\theta_m$ \\ (deg)\end{tabular} & \begin{tabular}[c]{@{}c@{}}nominal current \\ $I_m$ (A)\end{tabular} & \begin{tabular}[c]{@{}c@{}}turns circuit 1 \\ (2.5A)\end{tabular} & \begin{tabular}[c]{@{}c@{}}turns circuit 2 \\ (0.25A)\end{tabular} & \begin{tabular}[c]{@{}c@{}}turns circuit 3 \\ (0.025A)\end{tabular} & \begin{tabular}[c]{@{}c@{}}generated \\ current (A)\end{tabular} \\ \hline
\rowcolor[HTML]{EFEFEF} 
1  & 0.00  & -6.493  & -3  & 4  & 0   & -6.500  \\
2  & 20.41 & 4.630 & 2 & -2   & 5  & 4.625\\
\rowcolor[HTML]{EFEFEF} 
3  & 39.53 & -2.819  & -1   & -1   & -3   & -2.825\\
4  & 57.68 & 2.159 & 1  & -1   & -4   & 2.150\\
\rowcolor[HTML]{EFEFEF} 
5  & 75.46 & -1.662  & -1   & 3  & 4  & -1.650\\
6  & 93.03 & 1.478 & 1  & -4   & -1   & 1.475\\
\rowcolor[HTML]{EFEFEF} 
7  & 110.51 & -1.251 & 0    & -5   & 0  & -1.250\\
8  & 127.92 & 1.206 & 0   & 5  & -2  & 1.200\\
\rowcolor[HTML]{EFEFEF} 
9  & 145.30 & -1.080  & 0   & -4   & -3  & -1.075\\
10 & 162.65 & 1.101 & 0   & 4  & 4 & 1.100\\
\rowcolor[HTML]{EFEFEF} 
11 & 180.00 & -1.031  & 0   & -4   & -1  & -1.025\\
12 & 197.35 & 1.101 & 0   & 4  & 4 & 1.100 \\
\rowcolor[HTML]{EFEFEF} 
13 & 214.70 & -1.080  & 0   & -4   & -3 & -1.075\\
14 & 232.08 & 1.206 & 0   & 5  & -2 & 1.200\\
\rowcolor[HTML]{EFEFEF} 
15 & 249.49 & -1.251  & 0   & -5   & 0 & -1.250\\
16 & 266.97 & 1.478 & 0   & 6  & -1 & 1.475\\
\rowcolor[HTML]{EFEFEF} 
17 & 284.54 & -1.662  & -1   & 3  & 4 & -1.650\\
18 & 302.32 & 2.159 & 1  & -1   & -4  & 2.15\\
\rowcolor[HTML]{EFEFEF} 
19 & 320.47 & -2.819  & -1   & -1   & -3  & -2.825\\
20 & 339.59 & 4.630 & 2  & -2   & 5 & 4.625\\
\rowcolor[HTML]{EFEFEF} 
21 & - & -0.500 & - & - & - & -0.5        
\end{tabular}
\caption{Parameters for each of the wires forming the magnetic replicator. The angular position of each wire, $\theta_m$, follows the convention defined in Fig.~\ref{Figrender}c. Positive values of current indicate the current is flowing in the z-direction as defined in the figure. Wire $m=21$ corresponds to the central wire.}
\label{table1}
\end{table}

\subsection{Measurements} \label{sec.measurements}

The x and y components of the magnetic field were measured using two miniature fluxgate magnetometers (DRV425) whose sensor size is 4x4~mm and noise floor is around 1~nT/$\sqrt{\rm Hz}$. The two modules were placed one on top of the other, with a vertical gap of approximately 1~mm between them, and were oriented perpendicularly to measure the two components of the field. No cross-talk was measurable for the range of conditions presented.
A 3D-printed support structured held the two sensors at the right position. This structure was in turn attached to a plastic support arm which was mounted on two perpendicular manual linear stages. The error associated to the x,y coordinates of the sensors given by the linear stages was $\Delta x,y=0.1$mm. 

The two evaluation modules were connected to a 16-bit analogue to digital converter (Labjack T7) from which two voltages, $V_x$ and $V_y$, were obtained corresponding to the two sensors.


To perform a measurement of the field, we first recorded the background magnetic field for a period of 5 seconds at a sample rate of 132 Hz and calculated the mean value and the standard error of the measurements, $V_a^{\rm OFF}$ and $\Delta V_a^{\rm OFF}$, respectively ($a=\{x,y\}$). This was followed by measurements with the same settings in which the appropriate currents were on, obtaining $V_a^{\rm ON}$ and $\Delta V_a^{\rm ON}$. The magnetic field created by the currents is $V_a=V_a^{\rm ON}-V_a^{\rm OFF}$ and the associated standard error $\Delta V_a=\sqrt{(V_a^{\rm ON})^2+(V_a^{\rm OFF})^2}$. 
These two sets of recordings were repeated five times. To obtain a best estimate of the voltages, these five sets of values ($V_{a,i}$, $\Delta V_{a,i}$, $i=\{1,2,3,4,5\}$) were combined using the individual variances as weighting~\cite{Mandel2012}.

The sensors were calibrated in a known field, which was cross-checked with a precision fluxgate (Stefan Mayer Fluxmaster). The linearity was within 0.1\%.

\subsubsection{Numerical calculations}

In order to validate our measurements, we performed numerical calculations of the field created by different distributions of finite wires. We numerically integrated the 1D Biot-Savart equation to obtain the magnetic field at different positions of interest. The wires were assumed to be single uni-dimensional vertical current lines extending from $z=-200$mm to $z=200$mm (the connections between wires at the top and bottom ends of the structure were omitted). The current assigned to each wire corresponded to the "generated current" value in Table~\ref{table1}. 
The positioning of the wires in the xy plane was done by taking into account the measured position of the two reference points of the structure (see sect. \ref{sec.experimental}). The coordinates of these two points were used to find the center of the replicator and its relative rotation. Based on this information and on the nominal angles of the wires, $\theta_m$, we calculated the final x and y coordinates of each wire.

\subsubsection{Measurements of the field created by the magnetic replicator}

\begin{figure}[h!]
\centering
\includegraphics[width=0.95\textwidth]{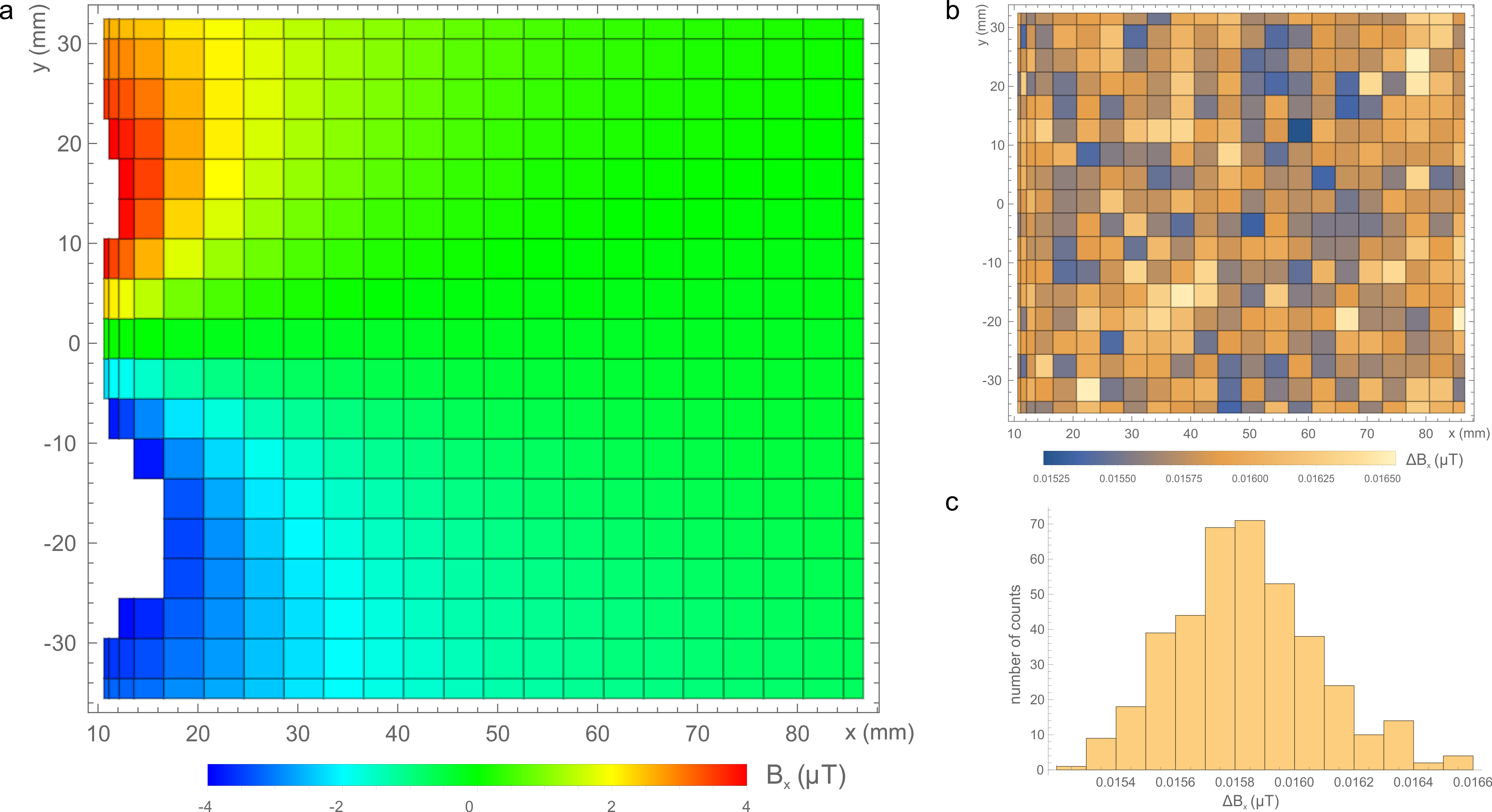} 
\caption{(a) Measured $B_x$. (b) Standard error corresponding to the measurements and (c) histogram of the errors.  \label{FigSemulationgridBx}}
\end{figure}

\begin{figure}[h!]
\centering
\includegraphics[width=0.95\textwidth]{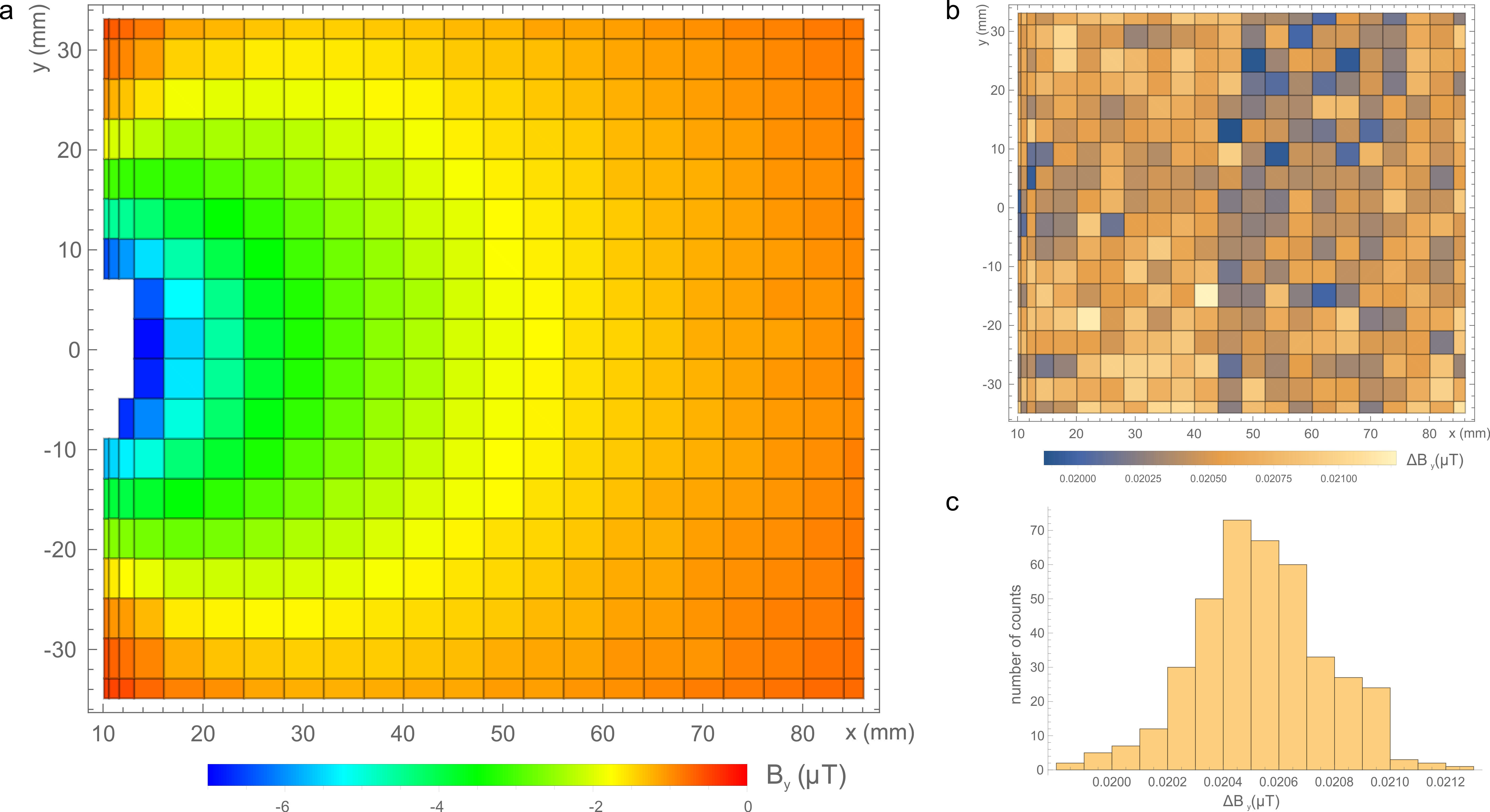} 
\caption{(a) Measured $B_y$. (b) Standard error corresponding to the measurements and (c) histogram of the errors.  \label{FigSemulationgridBy}}
\end{figure}

\begin{figure}[h!]
\centering
\includegraphics[width=1\textwidth]{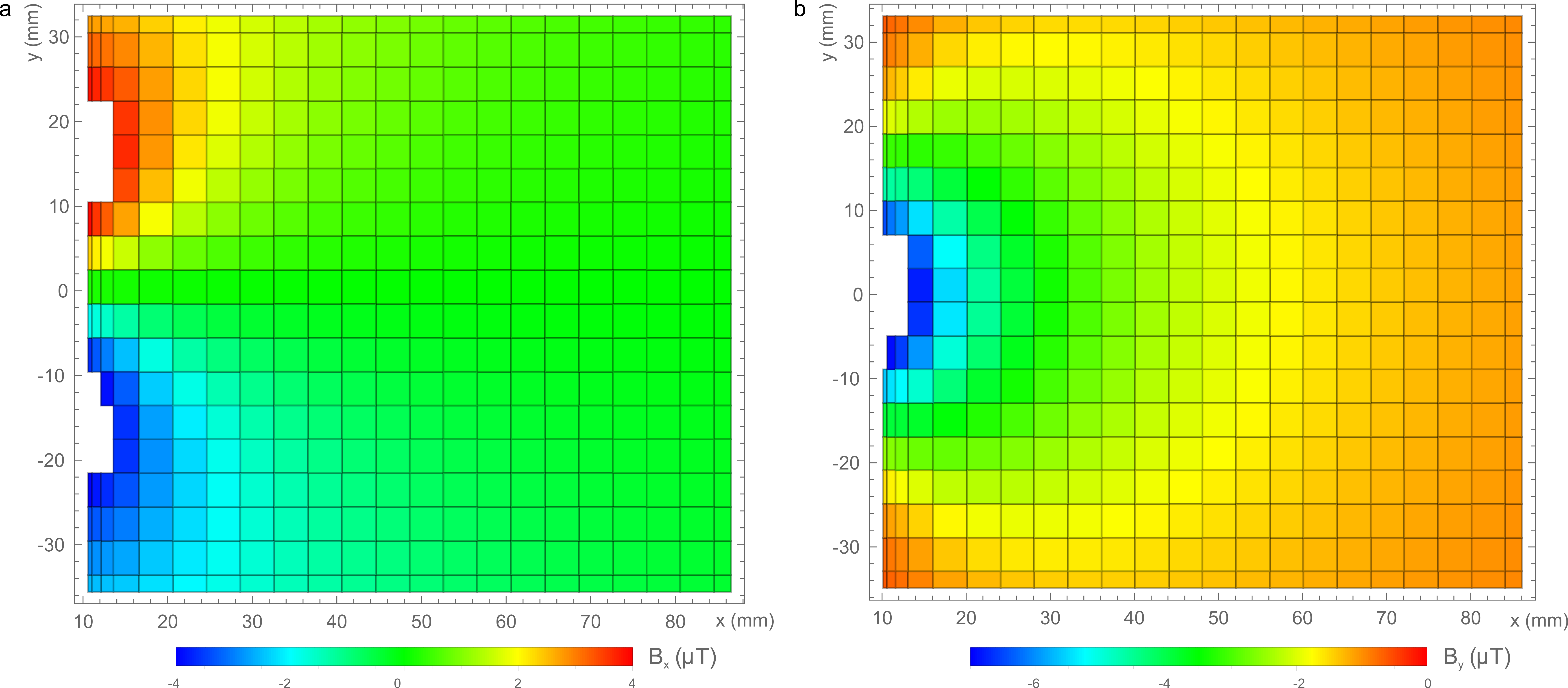} 
\caption{Numerical simulations of (a) $B_x$ and (b) $B_y$ obtained using Biot-Savart. \label{FigSemulationgrid}}
\end{figure}

We measured the magnetic field created by the magnetic replicator by feeding the 21 wires. The field was measured along the middle plane ($z=0$ in our coordinates system). Magnetic field values, $B_i$, and their corresponding standard errors, $\Delta B_i$ , were obtained as discussed in Sec.~\ref{sec.measurements}. Results are shown in Figs.~\ref{FigSemulationgridBx} and \ref{FigSemulationgridBy}. The corresponding numerical calculations are shown in Fig.~\ref{FigSemulationgrid}. We used the same color scales, ranges, and positions to make the grids directly comparable.

\begin{figure}[h!]
\centering
\includegraphics[width=1.0\textwidth]{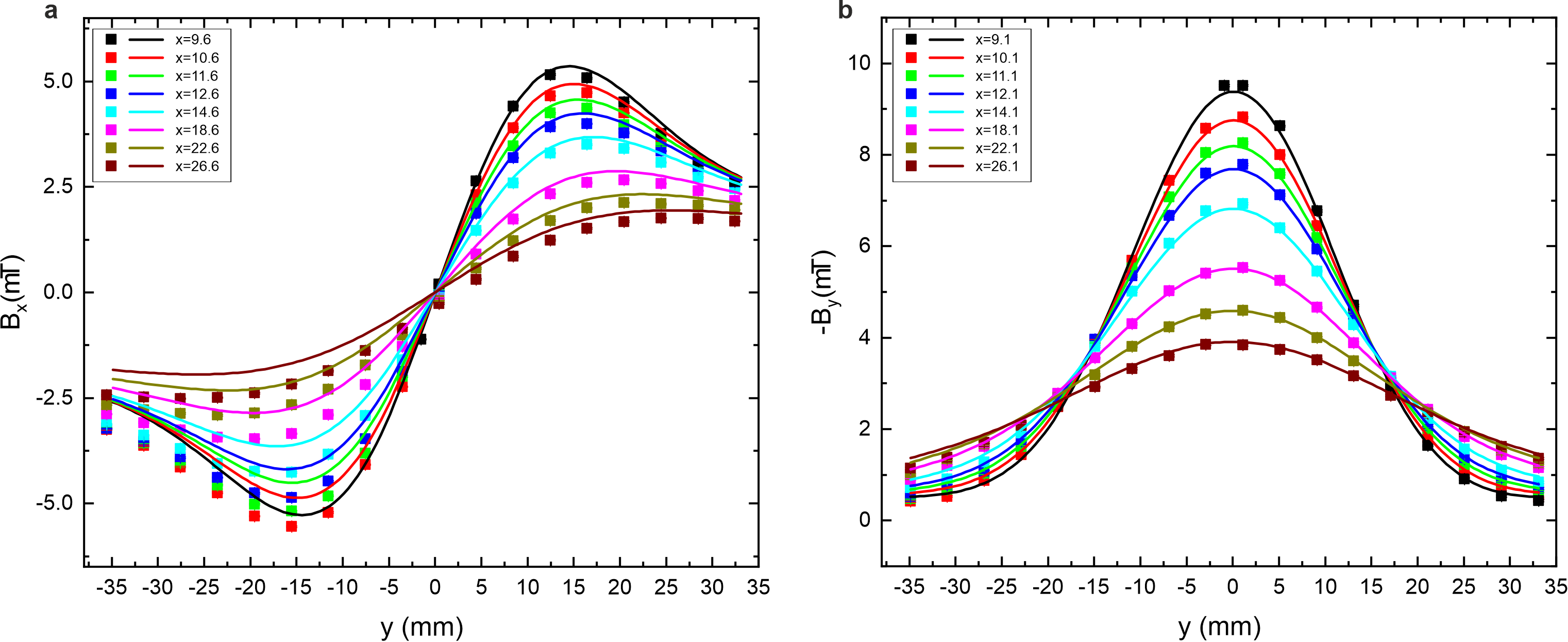} 
\caption{Measured $B_x$ (a) and $B_y$ (b) along lines of constant x (in symbols). Solid lines are the corresponding numerical calculations. \label{FigSemulationfocusing}}
\end{figure}

Figure \ref{FigSemulationfocusing} shows 1D plots of the measured fields (in symbols) and their corresponding calculations (solid lines). Fields are plotted as a function of y for different values of x corresponding to different parallel lines. 

\newpage
$\quad$
\newpage
$\quad$
\newpage
$\quad$

\subsubsection{Measurements of the cancellation of the field created by a wire}
\begin{figure}[h!]
\centering
\includegraphics[width=0.7\textwidth]{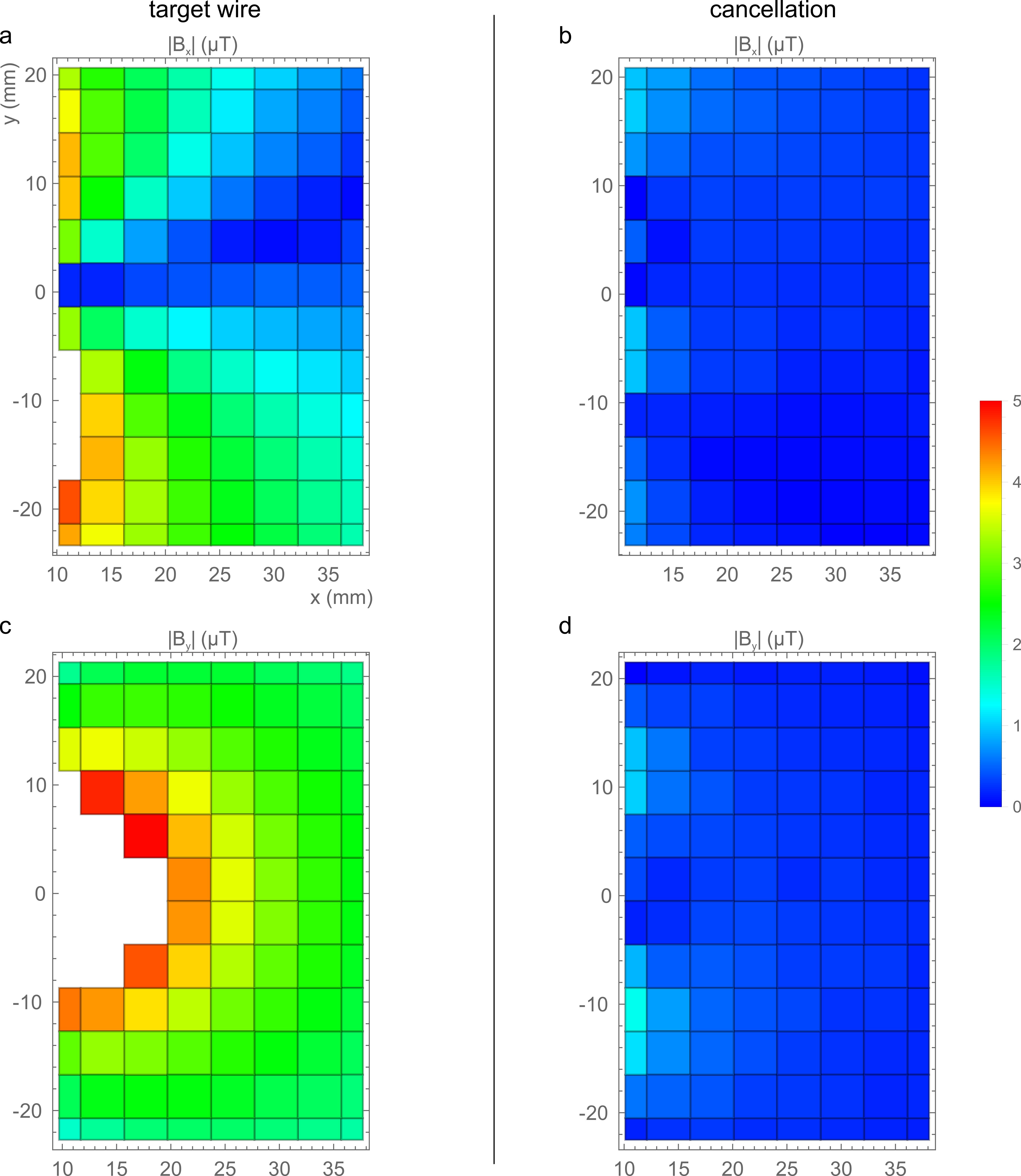} 
\caption{Measured field components when the target wire is fed [left, panels (a) and (c)] and when the metamaterial and the target wire are fed [right, panels (b) and (d)]. The largest standard error obtained in these measurements is $0.02\mu$T in all four panels.  \label{Figtargetvscancellation}}
\end{figure}

\begin{figure}[h!]
\centering
\includegraphics[width=0.5\textwidth]{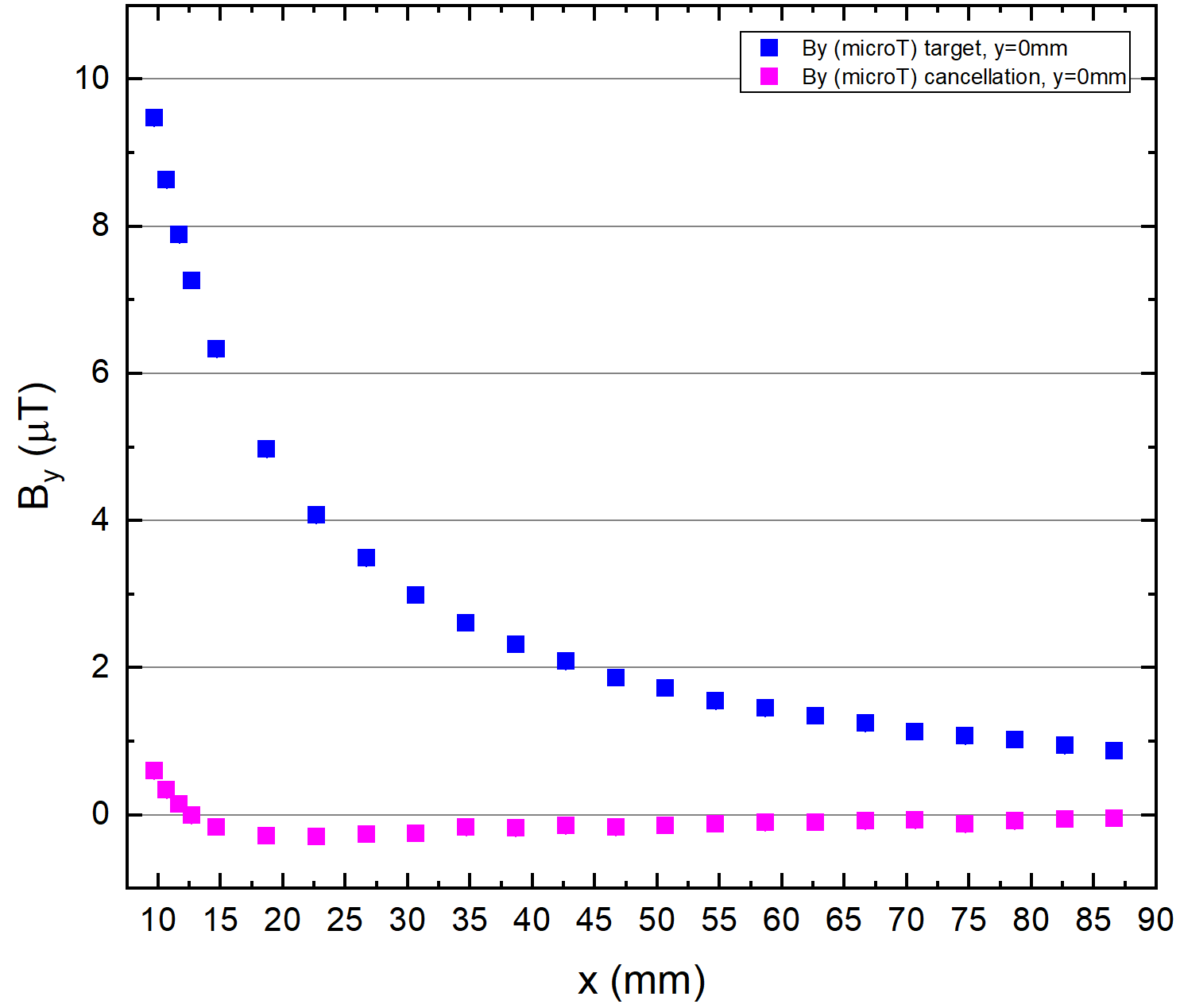} 
\caption{Measured field component $B_y$ when the target wire is fed (blue) and when the magnetic replicator and the target wire are fed (magenta).  \label{Figlinetargetvscancellation}}
\end{figure}

We measured the ability the magnetic replicator has to cancel the magnetic field created by a wire placed at the position of the replica wire. For this experiment we first measured the field created by a wire located at the position of the replica wire carrying $I=0.5$A. Results are shown on the left column of Fig.~\ref{Figtargetvscancellation} ("target wire"). We then fed the same wire together with the 21 wires forming the magnetic replicator. Results are shown on the right column of Fig.~\ref{Figtargetvscancellation} ("cancellation") and demonstrate a great reduction of the magnetic field. A 1D measurement along the $y=0$ line is shown in Fig.~\ref{Figlinetargetvscancellation}, confirming the cancellation is achieved up to large distances. These results show how the magnetic replicator can be used to cancel the field created by a long wire. 

\begin{figure}[h!]
\centering
\includegraphics[width=0.7\textwidth]{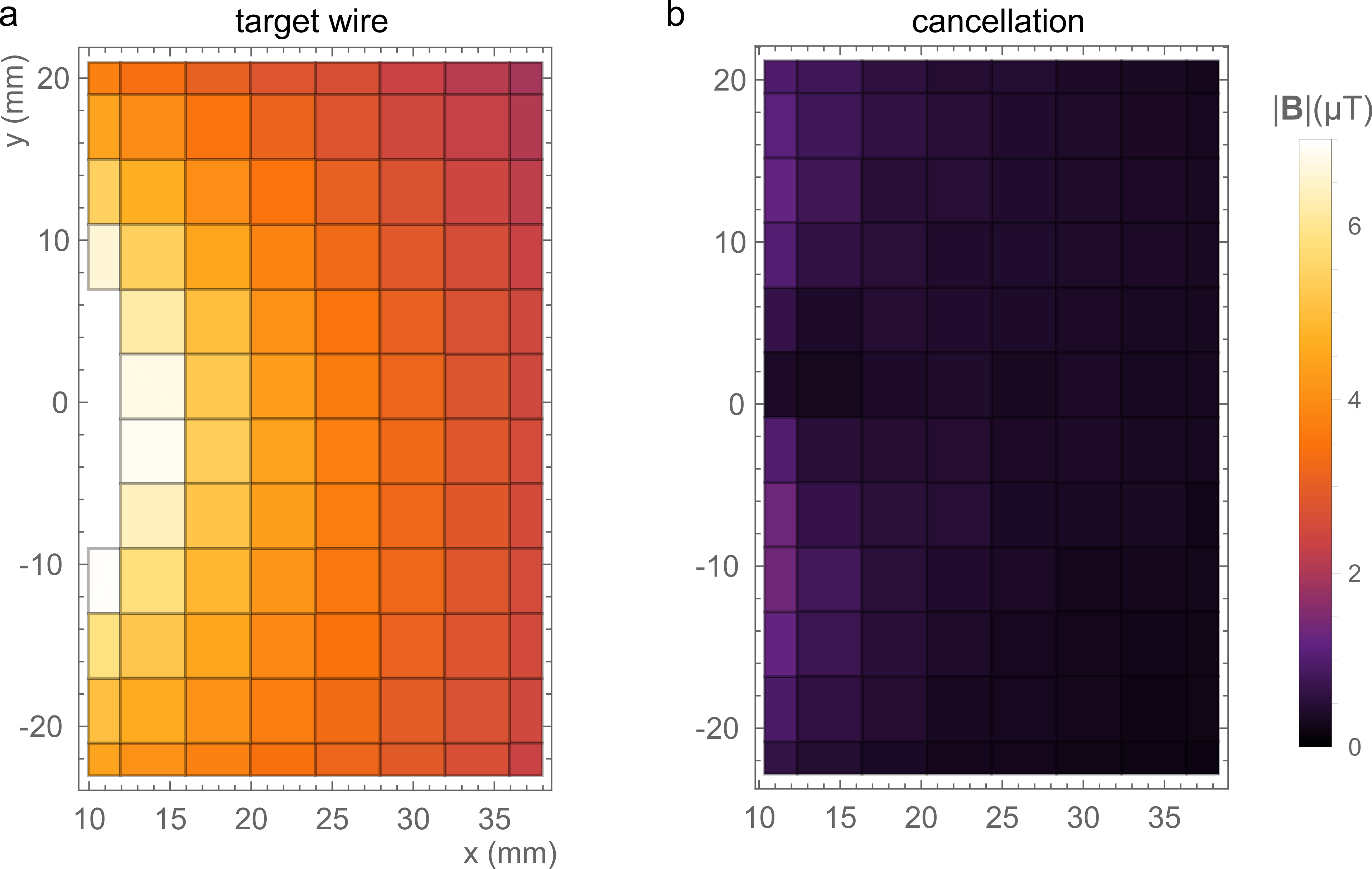} 
\caption{Measured absolute value of the field ($|\mathbf{B}|=\sqrt{B_x^2+B_y^2}$) when (a) the target wire is fed and (b) when the magnetic replicator and the target wire are fed. The largest standard error obtained in these measurements is $0.02\mu$T in the two panels. For each position, the coordinates corresponding to the $B_x$ and $B_y$ measurements where slightly different due to small differences in the position of the sensors. To generate this plot, we averaged the x and y coordinates. \label{FigtargetvscancellationAbs}}
\end{figure}

The two measured magnetic field components were combined to plot the absolute value of the magnetic field ($|\mathbf{B}|=\sqrt{B_x^2+B_y^2}$) for the "target wire" and "cancellation" cases. Results are shown in Fig.~\ref{FigtargetvscancellationAbs}.

\newpage
$\quad$
\newpage
$\quad$

\subsubsection{Measurements of the field created by the individual circuits}

\begin{figure}[h!]
\centering
\includegraphics[width=0.95\textwidth]{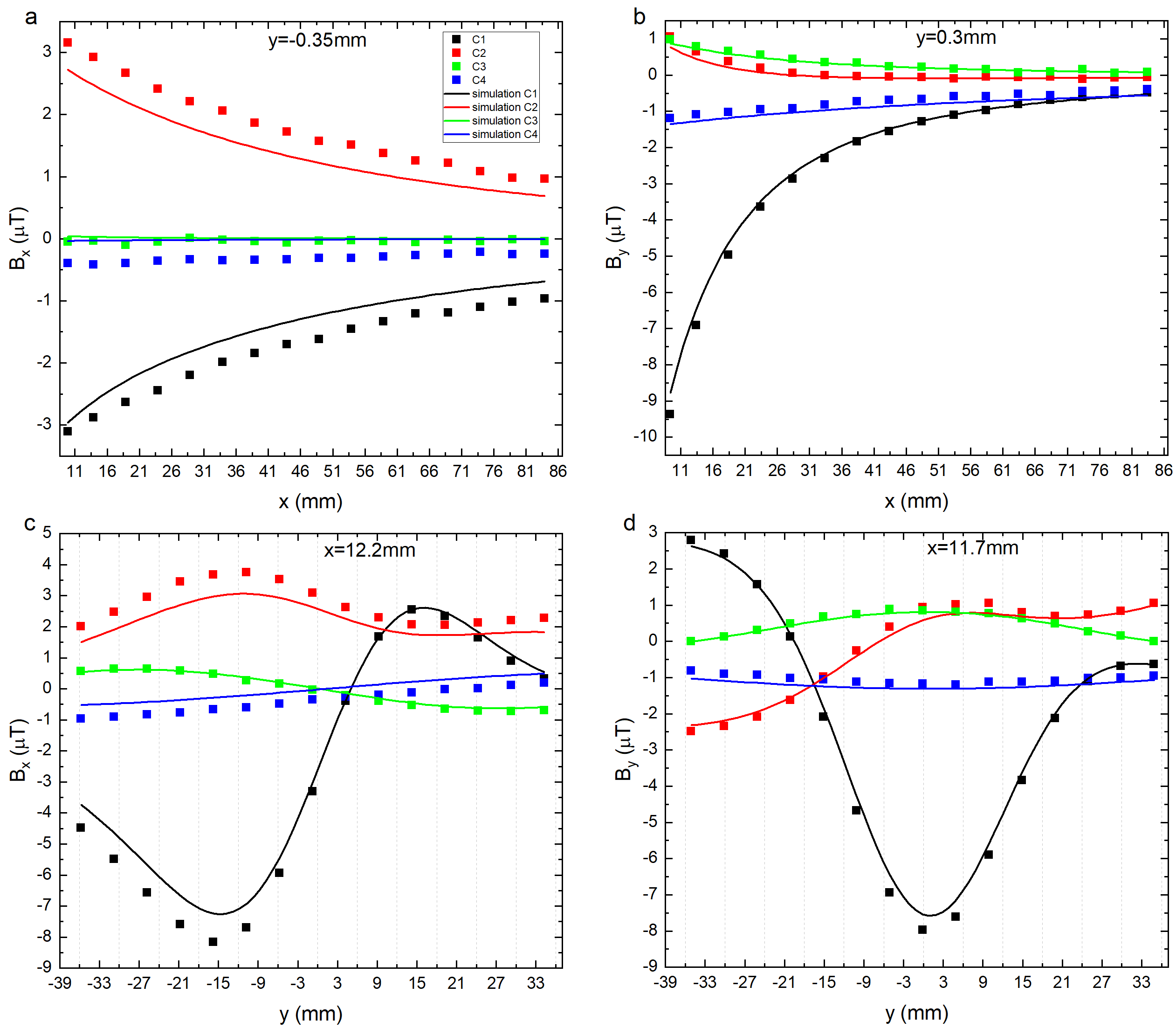} 
\caption{Measured (in symbols) and corresponding simulations of the fields created by the 4 independent circuits measured along the specified lines (the origin of corrdinates is set to the position of the target wire). The current supplied to the circuits was $I_{\rm C1}=2.5$A, $I_{\rm C2}=0.25$A, $I_{\rm C3}=0.05$A, $I_{\rm C4}=0.5$A.  \label{circuits}}
\end{figure}

We performed measurements of the field created by the 4 individual circuits forming the magnetic replicator. Results are shown in symbols together with their corresponding numerical calculations in Fig.~\ref{circuits}. $B_x$ and $B_y$ components were measured along two different lines. Since the two magnetic sensors measuring the different components where located at slightly different positions, the coordinates of the measuring lines for $B_x$ and $B_y$ do not match exactly. 

Measurements show that the 4 circuits create magnetic field distributions similar to the calculated values. The agreement is better for the y-components. 
To explain the potential sources of disagreement between measurements and calculations one first needs to consider that the calculations assumed a single straight wire located at the nominal position for each of the 21 wires forming the replicator. In the actual setup, several turns of wire were used in the 20 wire positions. The different circuits were winded one after the other, so that different turns were located at slightly different radial distances from the center of the structure. 
Another relevant source of error is the inaccurate verticality of the whole structure, which has a small base (a diameter of 80mm) compared to its height (of 400mm). 
Finally, the wires that supplied the current to the magnetic replicator could be responsible for creating undesired fields in the measuring area. The wires were not shielded (and we did not use coaxial wiring) and they were just moved as far as possible from the setup to minimize their effect. 

\newpage
$\quad$